\documentclass[useAMS,usenatbib]{mn2e}
\usepackage{graphicx,amsmath}

\makeatletter

\newcommand{\Rmnum}[1]{\expandafter\@slowromancap\romannumeral #1@}
\makeatother
\title[MHD instabilities in accretion mounds]{MHD instabilities in accretion mounds - \Rmnum{2}. 3D simulations}
\author[Mukherjee, Bhattacharya and Mignone]{Dipanjan Mukherjee$^{1}$\thanks{E-mail:
dipanjan@iucaa.ernet.in}, Dipankar Bhattacharya$^{1}$\footnotemark[1]\thanks{E-mail:
dipankar@iucaa.ernet.in} and Andrea Mignone$^{2}$ \thanks{E-mail: mignone@ph.unito.it }\\
$^{1}$Inter University Centre for Astronomy and Astrophysics, Post Bag 4, Pune 411007, India.\\
$^{2}$Dipartimento di Fisica Generale, Universita degli Studi di Torino
, Via Pietro Giuria 1, 10125 Torino, Italy}
\begin{document}

\date{In preparation}

\pagerange{\pageref{firstpage}--\pageref{lastpage}} \pubyear{2012}

\maketitle

\label{firstpage}

\begin{abstract}
We investigate the onset of pressure driven toroidal mode instabilities in accretion mounds on neutron stars by 3D MHD simulations using the PLUTO MHD code. Our results confirm that for mounds beyond a threshold mass, instabilities form finger like channels at the periphery, resulting in mass loss from the magnetically confined mound. Ring like mounds with hollow interior show the instabilities at the inner edge as well. We perform the simulations for mounds of different sizes to investigate the effect of the mound mass on the growth rate of the instabilities. We also investigate the effect of such instabilities on observables such as cyclotron resonant scattering features and timing properties of such systems.
\end{abstract}


\section{Introduction}
Neutron stars in binary systems accrete matter either from stellar winds \citep{ostriker73} or from Roche lobe overflow \citep{ghosh77,romanova02,romanova03} of the companion. The matter is channelled by the magnetic field to the poles, forming accretion mounds. Radiation from such mounds and the overlying columns gives rise to the pulsating X-ray emission observed due to the rotation of the neutron star (e.g. see \citet{patruno12} and \citet{caballero12} for some examples of accretion powered X-ray pulsars). 

The pressure of the confined accreted matter distorts the local magnetic field \citep{hameury83,melatos01,melatos04,dipanjan12} with important consequences for the  X-ray emission from the mounds and secular evolution of the neutron star's magnetic field. For high mass X-ray binary systems (hereafter HMXB) with companion masses  $\simeq 5-10 M_\odot$, the accreted matter confined within a polar cap of radius $\sim 1 $km generates strong broad band X-ray emission with characteristic cyclotron resonance scattering features (hereafter CRSF) (see \citet{coburn02,heindl04} for a review). Superposition of spectra from different parts of the mound with strong local distortions in the field lines would tend to make the CRSF structure broad and complex \citep{dipanjan12} (hereafter MB12). The observed broadening of CRSF at lower luminosity, e.g. in V0332+53 \citep{tsygankov10}, may be due to larger field distortions closer to the mound.

Over long term, the accreted matter would eventually flow horizontally along the neutron star surface. It has been proposed by various authors \citep{romani90,young95,cumming01,melatos01,melatos04} that such material flow may drag the field and bury it, leading to the relatively lower fields ($\sim 10^8-10^9$G) of neutron stars in low mass X-ray binaries (hereafter LMXB)\footnote{companion mass $\leq 1 M_\odot$} as compared to the $\sim 10^{12}$G fields in younger HMXB systems. However, plasma instabilities may disrupt such a burial processes by enabling cross-field transport of matter. Gravity driven modes may be triggered inside the accretion mound \citep{cumming01} resulting in the formation of closed disconnected loops \citep{melatos04,dipanjan12,dipanjan13a}. Such mounds are also prone to pressure driven instabilities \citep{litwin01}, as is common in systems like tokamaks where highly curved magnetic fields confine internal plasma (see \citet{friedberg82} for a review of MHD instabilities in such systems).

In this work we explore the MHD instabilities in accretion mounds strictly contained inside a polar cap of radius $\sim 1$km on a neutron star of surface field $\sim 10^{12}$G. This is appropriate for accretion mounds formed on HMXB and young LMXB with high magnetic field. The effects of gravity driven modes in such systems have been explored in \citet{dipanjan13a} (hereafter MBM13) via 2D axisymmetric simulations. In this paper we extend the work of MB12 and MBM13 to perform perturbation analysis in full 3D setting via MHD simulations.

This paper is organised as follows: i) in Sec.~\ref{Num.setup} we describe the equations solved to obtain the equilibrium structure of the mound which is subsequently perturbed and evolved to perform the MHD simulations, ii) in Sec.~\ref{Sec.results} we describe the results of the simulations for mounds of different sizes and structure. We show that higher mass mounds with large field distortions are highly unstable whereas mounds of smaller mass have slower growth rates tending to a stability threshold. iii) In Sec.~\ref{Sec.discuss} we discuss the implications of the MHD instabilities on a) the long term field evolution in such systems, b) the spectra and CRSF from mounds on HMXB systems and c) the timing features expected from such mounds.

\section{Numerical set up}\label{Num.setup}
The analytical and numerical techniques employed to obtain the equilibrium solution of the mound structure have already been discussed in detail in MB12 and MBM13. In this section we briefly summarise the steps and discuss the differences between the 2D and 3D simulations. Following MB12, we first solve for the equilibrium field configuration of the static mound by solving a Grad-Shafranov (hereafter GS) equation in cylindrical coordinate system:
\begin{equation}\label{GSeq}
\frac{\Delta ^2 \psi}{4\pi r^2} = -\rho g \frac{dZ_0}{d\psi}
\end{equation}
where the GS operator is defined as: $\Delta ^2 = r\frac{\partial}{\partial r}(\frac{1}{r}\frac{\partial }{\partial r}) + \frac{\partial ^2}{\partial z^2}$. The flux function $\psi$ is related to the poloidal magnetic field as $\mathbf{B} = (\boldsymbol{\nabla} \psi \times \hat{\boldsymbol{\theta}})/r$. We assume zero toroidal magnetic field. We consider a degenerate non-relativistic plasma (with $\mu _e =2$) whose pressure is given by
\begin{equation}\label{eq.eos}
p=3.122\times 10^{12} \rho^{5/3} \mbox{ dyne cm}^{-2}
\end{equation}
$\rho$ being the density in g ${\rm cm}^{-3}$. We assume Newtonian gravity with constant acceleration $g=1.86\times 10^{14} \mbox{cm s}^{-2}$ for a neutron star of radius $10$km and mass $1.4 M_{\odot}$. As in MBM13, we use a GS solution for a polar field $\mathbf{B}_{\rm p}=B_0 \hat{\mathbf{z}}$, with $B_0=10^{12}$G at the surface. The shape of the mound is defined by the mound height function $Z_0(\psi)$ (see MB12 for more on the mound height function). Following MBM13, for a filled mound, we consider the profile
\begin{equation}\label{parabolicpro}
Z_0(\psi ) = Z_{\rm c}\left(1-\tilde{\psi}^2\right) 
\end{equation} 
and for a hollow mound we use
\begin{equation}\label{holopro}
Z_0(\psi ) = \frac{Z_{\rm c}}{0.25}\left(0.25-\left(\tilde{\psi} - 0.5\right)^2\right)
\end{equation}
where $\tilde{\psi}=\psi/\psi_p$, $\psi _p = 1/2(B_0 R^2_p)$ is the flux function at the polar cap radius $R_p=1$km. The apex height is given by $Z_c$ which for a filled mound is at $r=0$ and for a hollow mound at $r(\tilde{\psi}=0.5)$.

The equilibrium solutions from the GS solver are used as initial conditions in the PLUTO MHD code \citep{andrea07}. The PLUTO code solves the full set of ideal MHD equations following the Godunov scheme (e.g. MBM13, \citet{andrea07}). As the density falls off to zero beyond the mound height, we choose a section of the GS solution as the PLUTO domain (as in MBM13), such that local Alfv\'en speeds are non-relativistic. The 2D axisymmetric GS solutions carried out over a polar plane ($r-z$) are rotated in the azimuthal direction to construct the 3D mound in force equilibrium. The GS solutions are then imported into PLUTO by a trilinear interpolation scheme. The simulation is carried out in $\left[0,\pi/2\right]$ domain in the azimuthal direction with periodic boundary condition at both ends. Although this does impose an assumed symmetry of a quarter of a quadrant, the results are qualitatively similar to that of a full azimuthal domain $[0,2\pi]$, as discussed in Sec.~\ref{Sec.toroidal_modes}. 

 To minimise numerical errors, the grid spacing was chosen such that the grids are nearly cubic at the centre of the PLUTO domain ($r_c\Delta \theta \simeq \Delta r \simeq \Delta z$, where $r_c$ is midway between the radial extremes of the PLUTO domain). The runs were performed with different grid resolutions to check for the convergence of the solutions (see Sec.~\ref{Sec.toroidal_modes} for details on comparisons of resolutions). It was found that choosing spatial resolutions less than a metre was sufficient to follow the growth of the instabilities and their subsequent effects (see Table~\ref{tableres} for some typical resolutions used).  

\begin{table}
\centering
\begin{tabular}{l l l l}
\hline
$Z_c$ & $N_r \times N_\theta \times N_z$ & $\Delta l$ ($\Delta r \simeq r_c \Delta \theta \simeq \Delta z $) \\
\hline
{\bf Solid mound} & & \\
70m  & $1024\times 1536 \times 96$ & $\sim 0.59$m \\
50m  & $744 \times 1168 \times 48$ & $\sim 0.67$m \\
45m  & $824 \times 1512 \times 48$ & $\sim 0.55$m \\
{\bf Hollow mound} & & \\
45m  & $572 \times 2096 \times 64$ & $\sim 0.52$m \\
\hline
\end{tabular}
\caption{Sample resolutions for simulation runs.}
\label{tableres}
\end{table}

Normalised density and magnetic field (see Table~\ref{tablevar} for normalisation constants) are interpolated from the GS solutions.  Pressure is evaluated from eq.~(\ref{eq.eos}).  Initial velocities and toroidal field are set to zero. For our simulations we have used the HLL Riemann solver \citep{toro08}, a third order Runge-Kutta scheme for time evolution and a third order piece-wise parabolic scheme for interpolation (PPM as in \citet{colella84}). For preserving the $\nabla\cdot\boldsymbol{B}=0$ constraint, we use the extended generalised Lagrange multiplier scheme (EGLM) (e.g. \citet{dedner02}, \citet{mignone10_1}, \citet{mignone10_2}) as it gives improved numerical stability. See MBM13 for more discussion on the relevant schemes which are suitable for the problem undertaken. 

At the base of the mound ($z=0$) we apply a fixed boundary condition to represent a hard crust. At the top and the sides we apply a fixed gradient boundary condition, where the gradients of the physical parameters are kept fixed to the initial value.  This preserves the force equilibrium at the domain boundaries, preventing any artificial gradients that may lead to numerical errors. Such a boundary condition signifies an outflow on the perturbed quantities (see MBM13).
\begin{table}
\centering
\begin{tabular}{l l}
\hline
Parameters & values \\ \hline
Magnetic field ($B_p$) & $10^{12}$G \\ %\hline
Density ($\rho _0$) & $10^6 \mbox{g cm}^{-3}$ \\ %\hline
Length ($L_0$) & $10^5$ cm \\ %\hline
Velocity ($V_{A0} = B_p/\sqrt{4 \pi \rho_0}$) & $2.82 \times 10^8 \mbox{cm s}^{-1}$ \\ %\hline
Time ($t_A = L_0/V_{A0}$) & $3.55 \times 10^{-4}$s \\
\hline
\end{tabular}
\caption{Table of normalisation constants for a simulation with GS solution of a mound with polar field $B_p = 10^{12}$G.}
\label{tablevar}
\end{table}

The equilibrium GS solution is first evolved in PLUTO without any added perturbation to test the stability of the numerical schemes. For the set of schemes used, the equilibrium solution remains intact without any substantial internal flows. The maximum internal flow velocities at $t\sim 2 t_A$ for a 70m mound with resolution $512\times768\times48$ is less than 1\% of local Alfv\'en and magnetosonic velocities. 
\begin{figure}
	\centering
        \includegraphics[width = 7.cm, height = 7.cm,keepaspectratio] {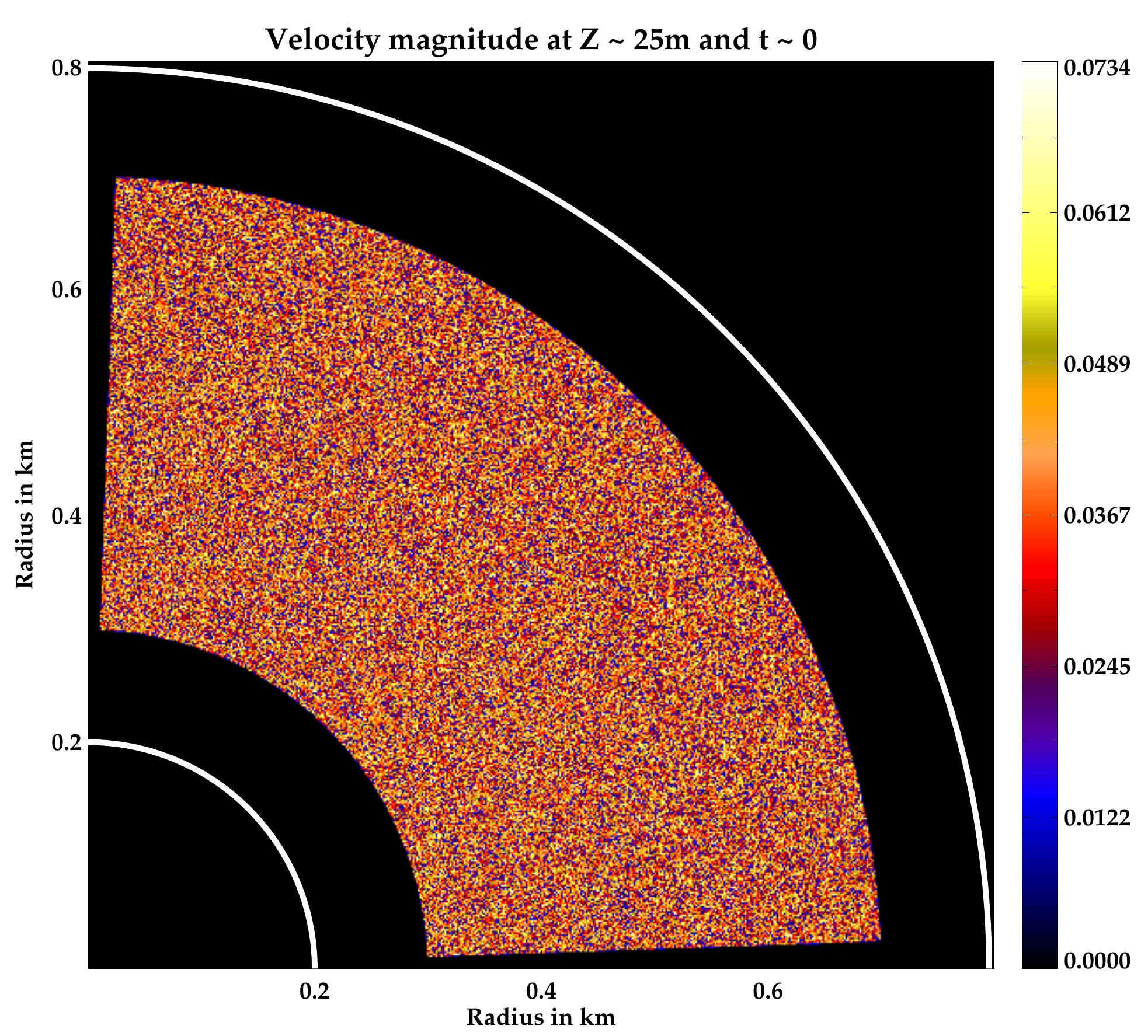}
	\caption{\small A horizontal section at $z \sim 25$m of a mound of height $\sim 70$m, showing the initial normalised random velocity field (see Table~\ref{tablevar} for normalisation constants). The white lines denote the radial edges of the PLUTO domain in a cylindrical coordinate system ($r,\theta,z$). The simulation is set up for azimuthal range $[0,\pi /2]$. } \label{velocity_pl0} 
\end{figure}

The system is then perturbed by adding a random velocity field to the static equilibrium GS solutions (see Fig.~\ref{velocity_pl0}) with the maximum strength set to a fraction of the squared mean of the local Alfv\'en speed and sound speed:
\begin{equation}\label{perteq}
v_{(r,\theta,z)}= \xi (r,\theta,z) \sqrt{c_s^2 + v_A^2}
\end{equation}
where $v_{(r,\theta,z)}$ corresponds to any of the three velocity components, $\xi$ is a random number in the range ($-\eta,\eta$) assigned to a grid point, $\eta$ giving the maximum strength of the local perturbation, $c_s^2=(5/3)p/\rho$ and $v_A^2=B^2/(4\pi\rho)$. The perturbation strength is fixed to a small value to excite linear modes ($\eta \sim 2\times 10^{-2}$).

\section{Results}\label{Sec.results}
\subsection{Toroidal mode instabilities in filled mounds}\label{Sec.toroidal_modes}
\begin{figure*}
	\centering
        \includegraphics[width = 5.8cm, height = 5.5cm] {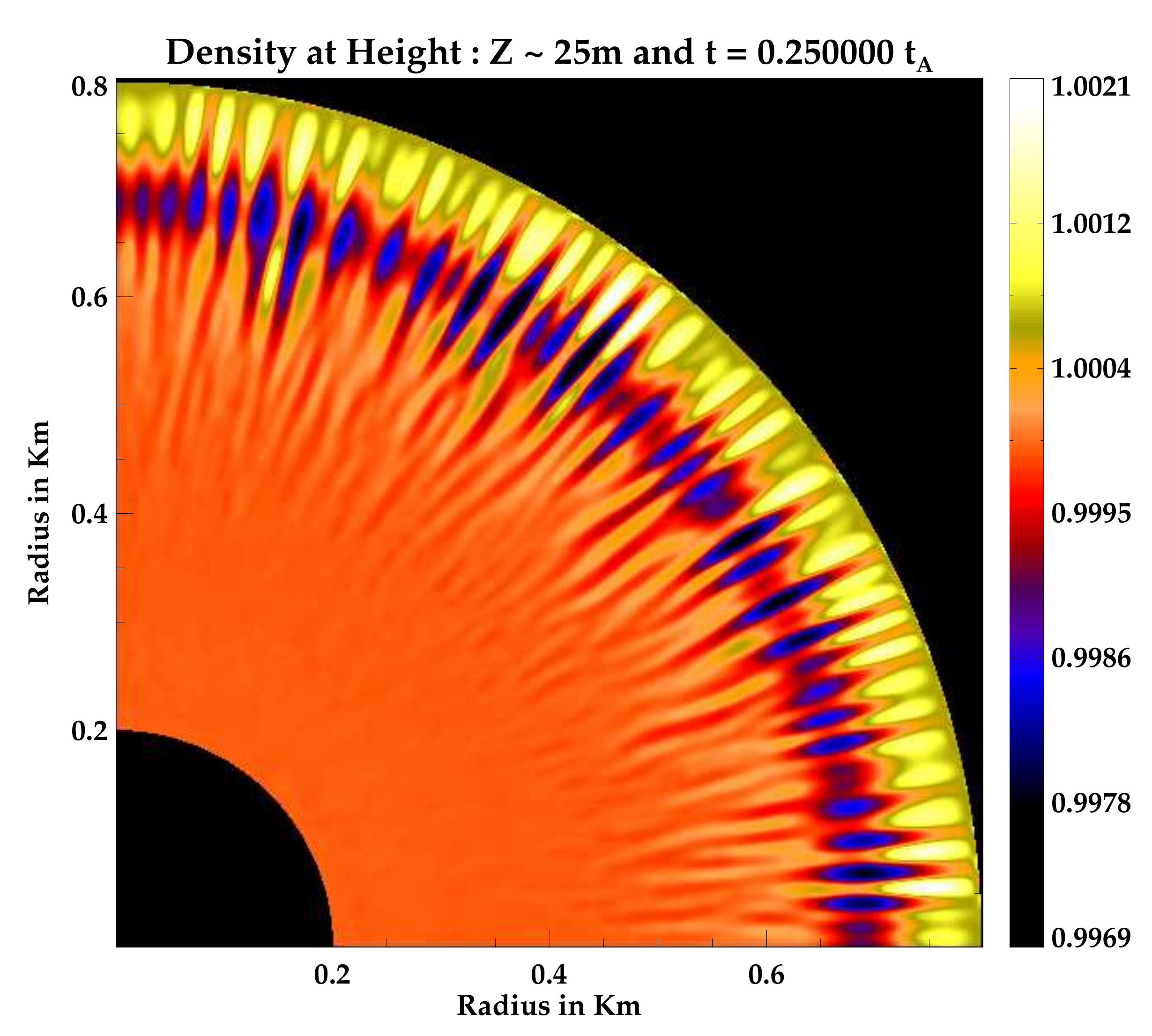}
	\includegraphics[width = 5.8cm, height = 5.5cm] {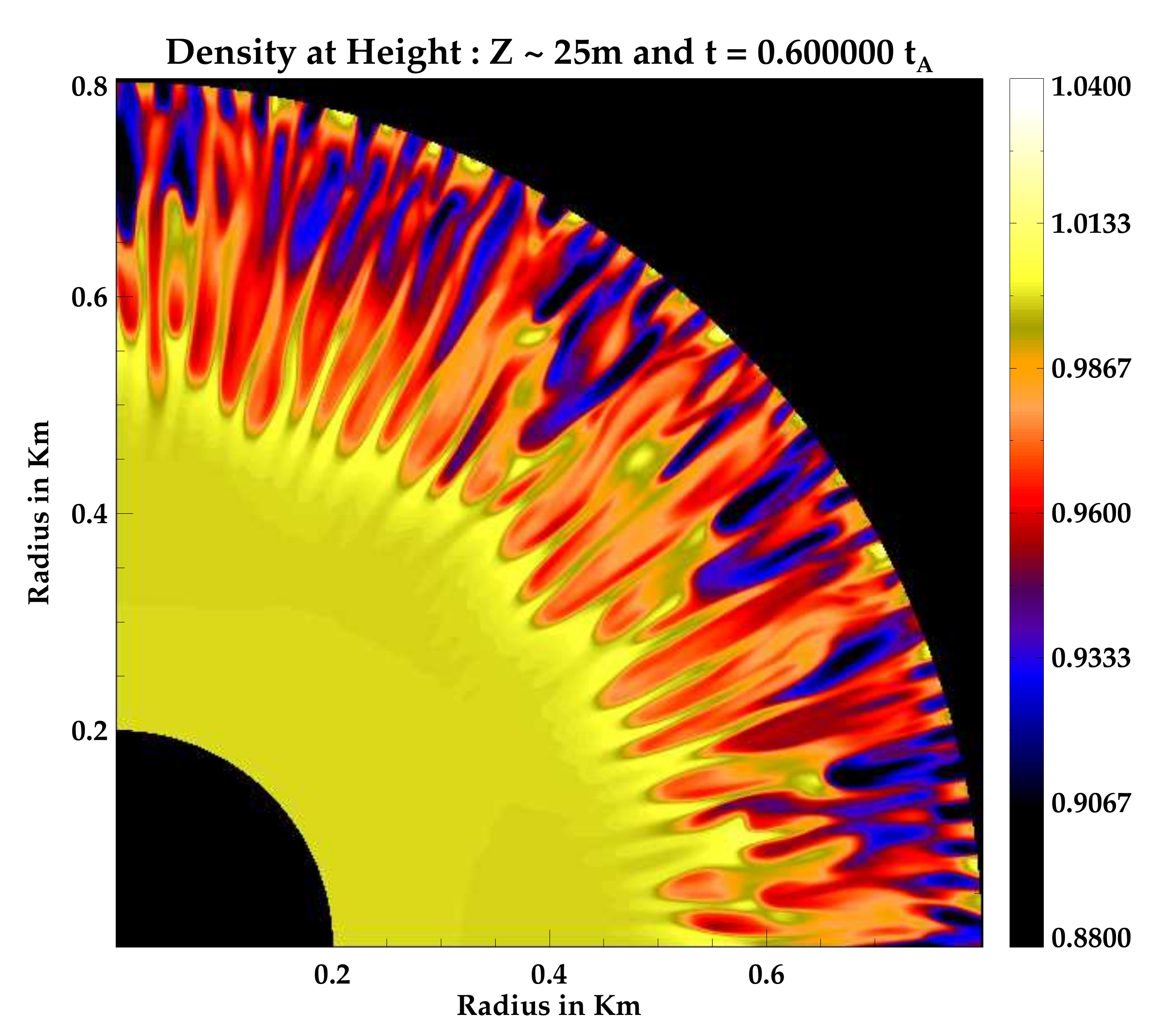}
	\includegraphics[width = 5.8cm, height = 5.5cm] {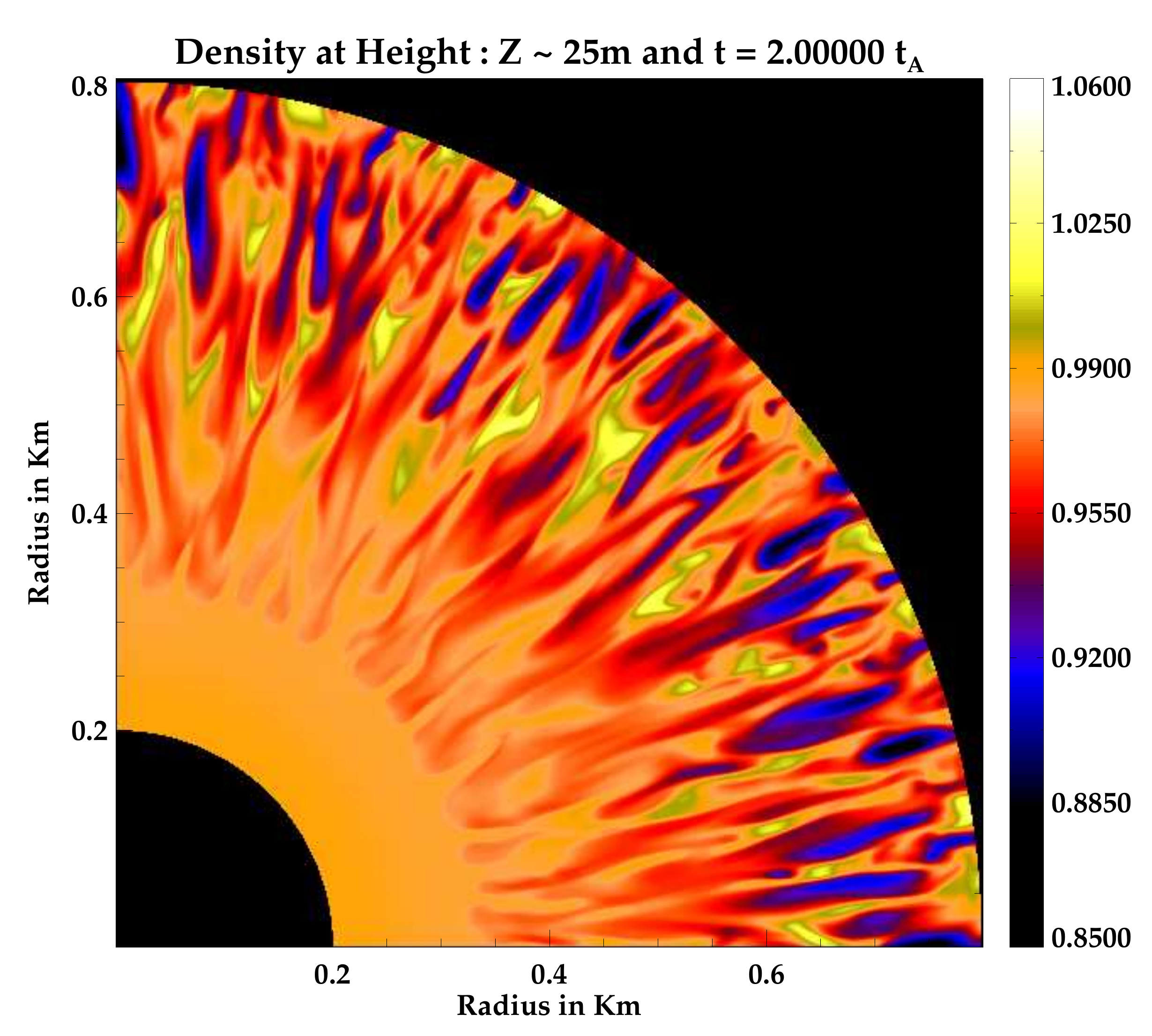}
	\caption{ \small Density normalised to equilibrium value showing the development of the MHD instabilities at different times. At the onset of the instabilities radial fingers of over-dense structures are formed due to development of toroidal modes. With time the radial fingers merge and spread throughout the mound. }\label{density_slices_70m}
\end{figure*}
\begin{figure}
	\centering
        \includegraphics[width = 10.cm, height = 9.cm,keepaspectratio] {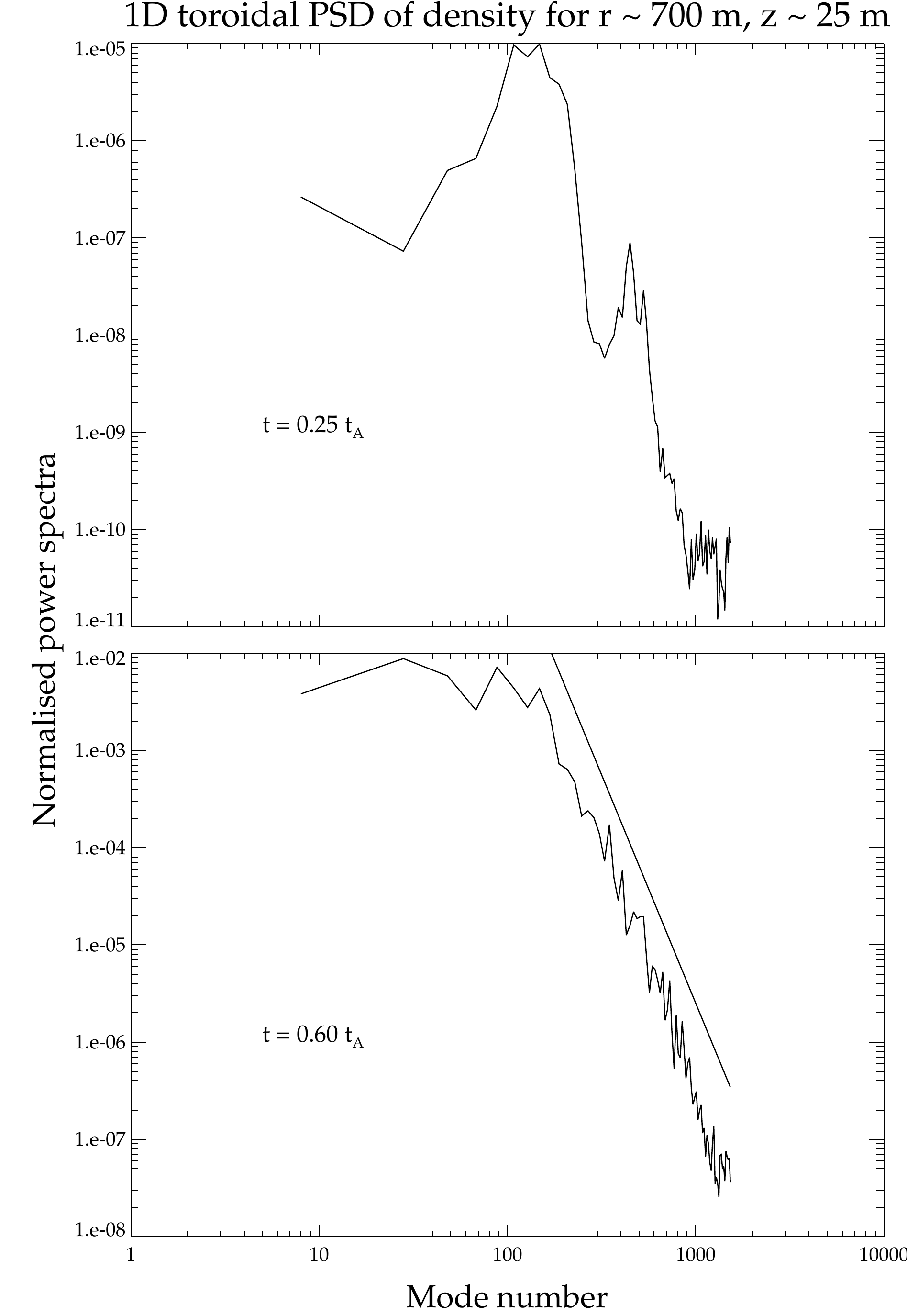}
	\caption{\small 1D power spectral density at $r\sim700$m and $z\sim25$m at times $t \sim 0.2 t_A$ and $\sim 0.6 t_A$ respectively (corresponding to the first two panels in Fig.~\ref{density_slices_70m}). The x axis gives the mode number (see text). A line is drawn parallel to the PSD in the second plot to show the powerlaw nature of the Fourier spectra.} \label{1Dpsd}
\end{figure}
\begin{figure}
	\centering
        \includegraphics[width = 7.cm, height = 7.cm,keepaspectratio] {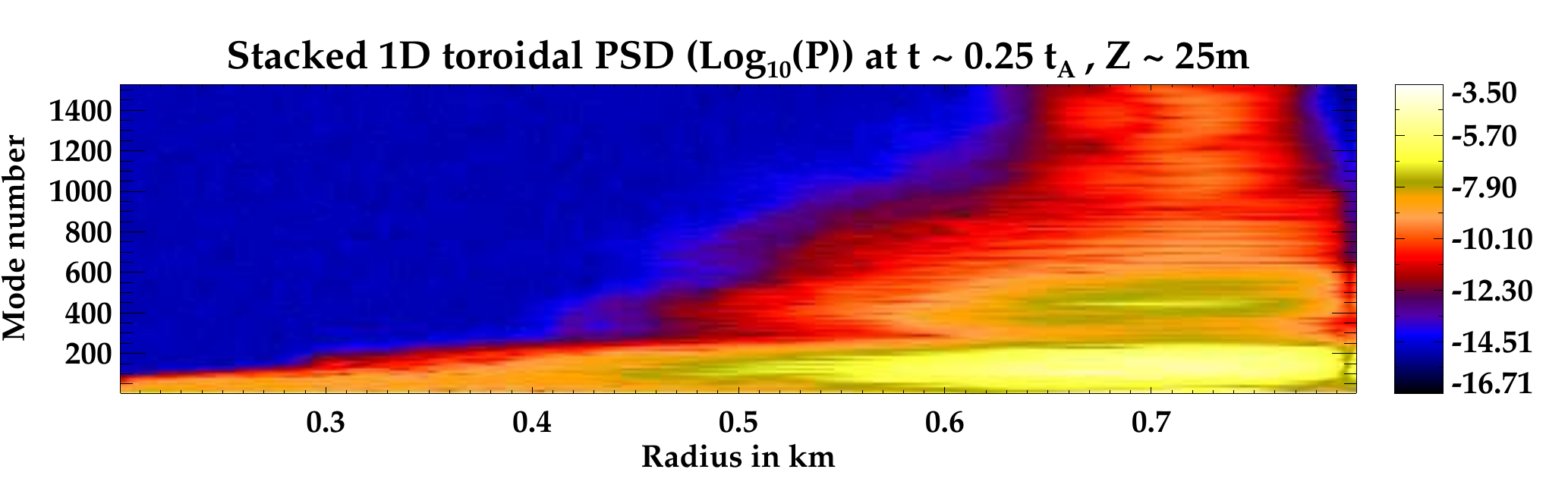}
        \includegraphics[width = 7.cm, height = 7.cm,keepaspectratio] {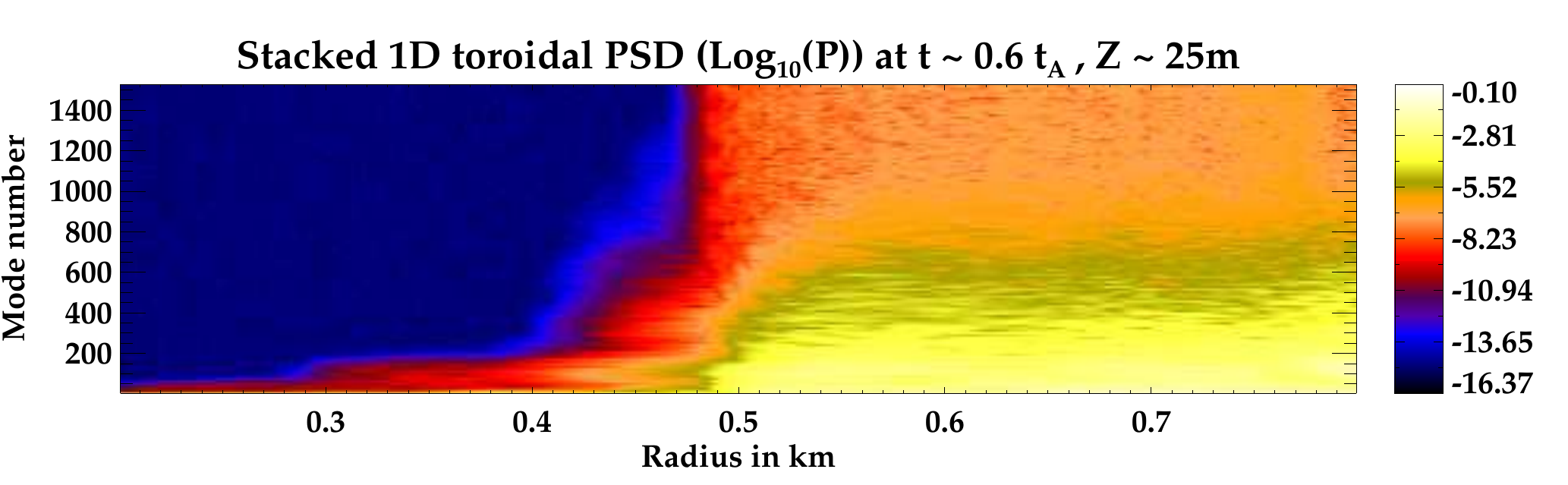}
        \includegraphics[width = 7.cm, height = 7.cm,keepaspectratio] {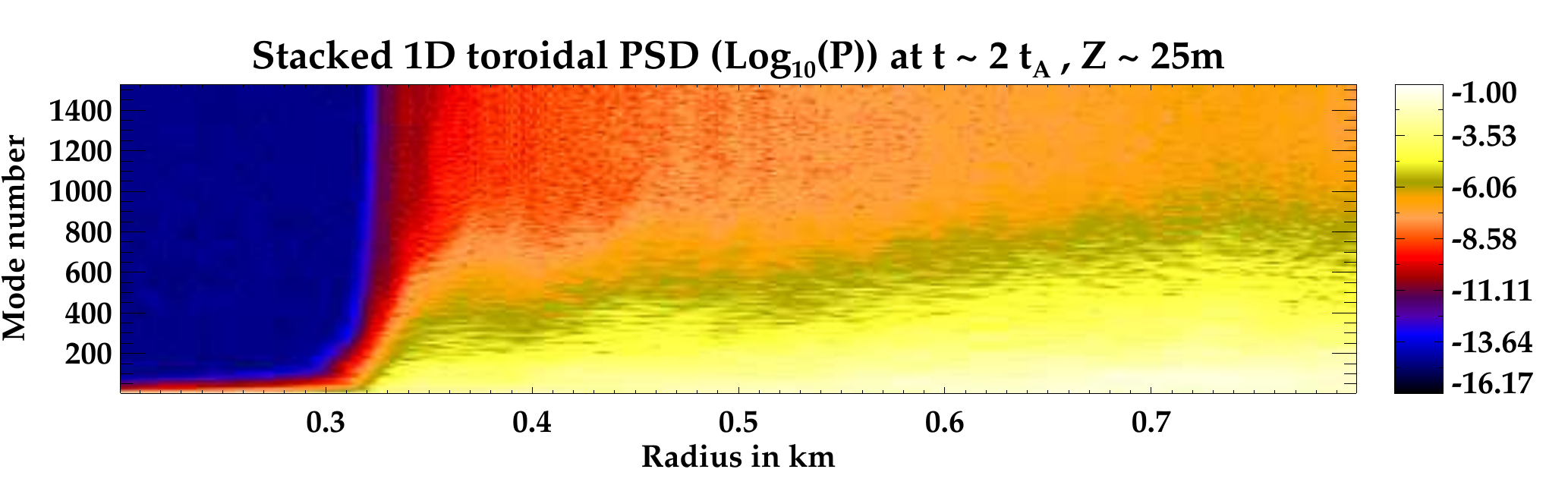}
	\caption{\small Stacked 1D power spectral density of perturbed density at different radii and times (corresponding to the panels in Fig.~\ref{density_slices_70m}). At the initial stages of the development of the instabilities ($t\sim0.25 t_A$), power is concentrated in some distinct modes. At later times, power is spread over all scales. The instability is seen to spread inwards as there is increase in power at the inner radii with time. } \label{2Dpsd}
\end{figure}
{\bf Non-axisymmetric structures: } For a mound near the GS threshold ($Z_c = 70$m following eq.~(\ref{parabolicpro}), with $B_p=10^{12}$G), applied perturbations easily excite MHD instabilities resulting in the growth of toroidal modes. At the onset of the instability, several radial finger like channels are formed which later merge to form random over and under-dense regions (see Fig.~\ref{density_slices_70m}). To investigate the structures developed by the toroidal modes we perform one dimensional discrete Fourier transform along the azimuthal direction at certain fixed values of time, radius and height: 
\begin{equation}\label{eqFT}
\tilde{Q}(u_n)=\frac{1}{N_\theta}\sum _{l=0} ^{N_\theta -1} Q(\phi_l) e^{-i2\pi u_n\phi_l}
\end{equation}
where $Q$ is any physical parameter (e.g. density) and $\tilde{Q}$ is its Fourier transform, $\phi_l=l\pi/(2N_\theta) $ is the azimuthal coordinate and $N_\theta$ is the total number of cells in the azimuthal direction. The corresponding frequencies in the Fourier domain are $u_n = n/(\pi/2)$. Using the above in eq.~(\ref{eqFT}), we get the power spectral density (PSD) as 
\begin{equation}\label{eqDFT}
P(u_n)= \left | \frac{1}{N_\theta}\sum _{l=0} ^{N_\theta-1} Q(\phi_l) e^{-i4n\phi_l} \right | ^2
\end{equation}
where $m=4n$ can be identified as the mode number for Fourier modes over the whole domain of $[0,2\pi]$ ($\sim \exp(-im\phi)$). The one dimensional PSD of perturbed density at $r\sim 700$m and $z\sim25$m for two different times are presented in Fig.~\ref{1Dpsd}. The 1D PSD at different radii are stacked and represented as a colour contour plot in Fig.~\ref{2Dpsd}. At initial stages of the development of the instabilities ($t\sim 0.25 t_A$), power is concentrated in small wavelength modes in the outer half of the mound, as alternating finger like channels are produced. As the instabilities saturate, they proceed inwards into the mound and power is spread over all scales in a power-law fashion. 

\begin{figure*}
	\centering
        \includegraphics[width = 8.5cm, height = 8.5cm,keepaspectratio] {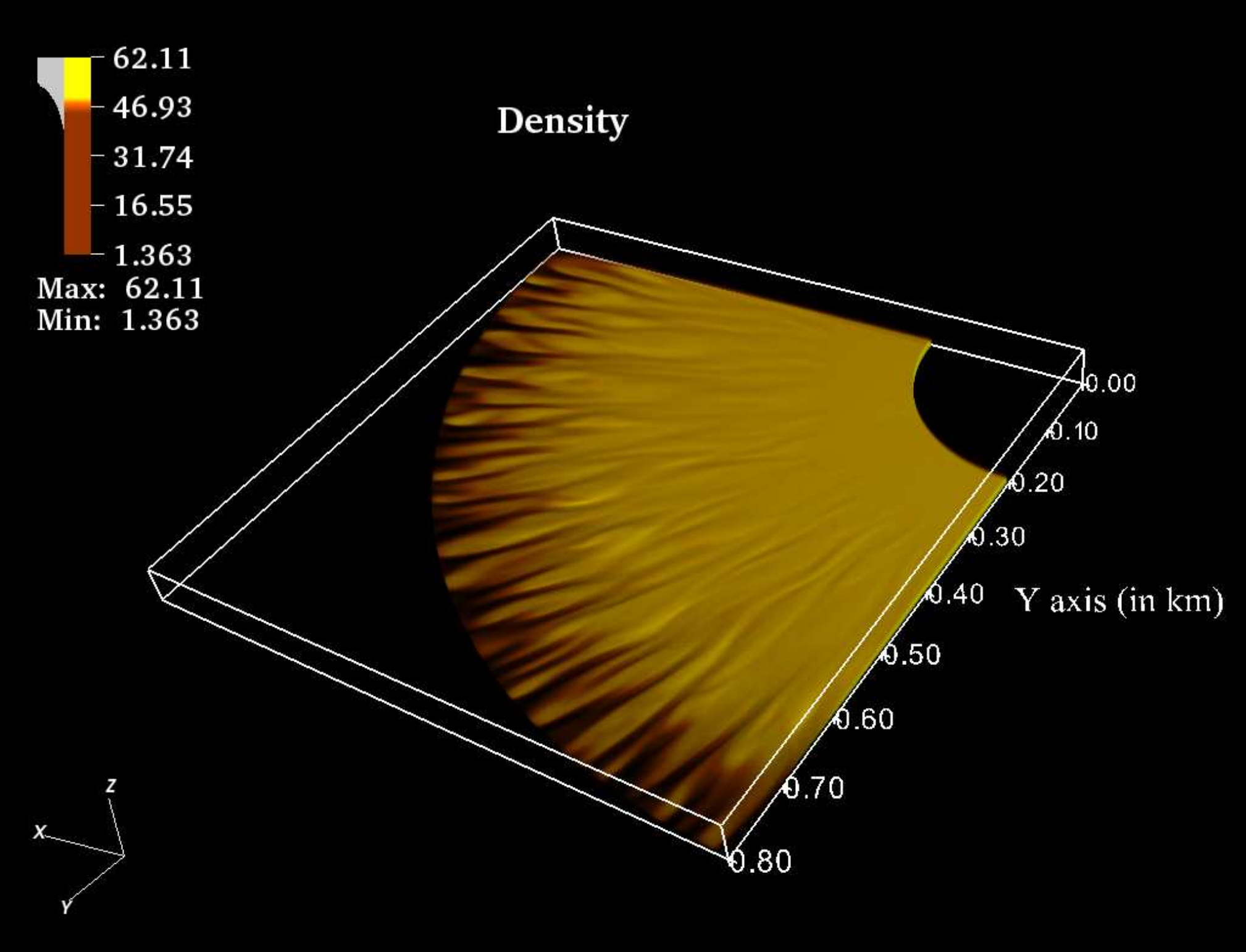}
	\includegraphics[width = 8.5cm, height = 8.5cm,keepaspectratio] {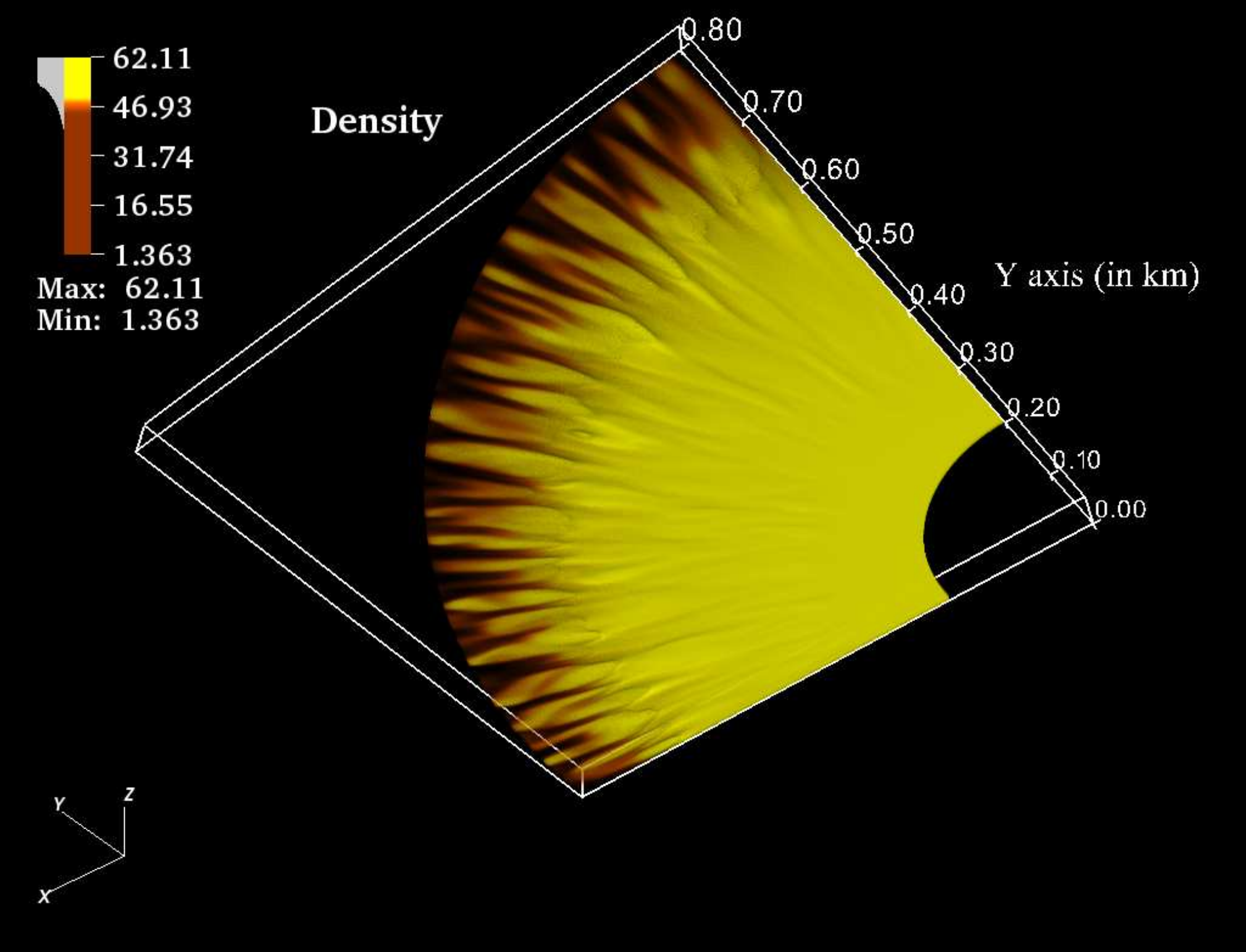}
	\caption{\small The top (left figure) and bottom (right figure) view of 3D density contours at $t\sim 2t_A$ for $Z_c=70$m filled mound with $B_p=10^{12}$G.  The density is in the units of $10^6 \mbox{ g cm}^{-3}$. The finger like channels are clearly seen at the outer edge.}\label{density70m}
\end{figure*}
\begin{figure*}
	\centering
        \includegraphics[width = 8.5cm, height = 6.64169cm] {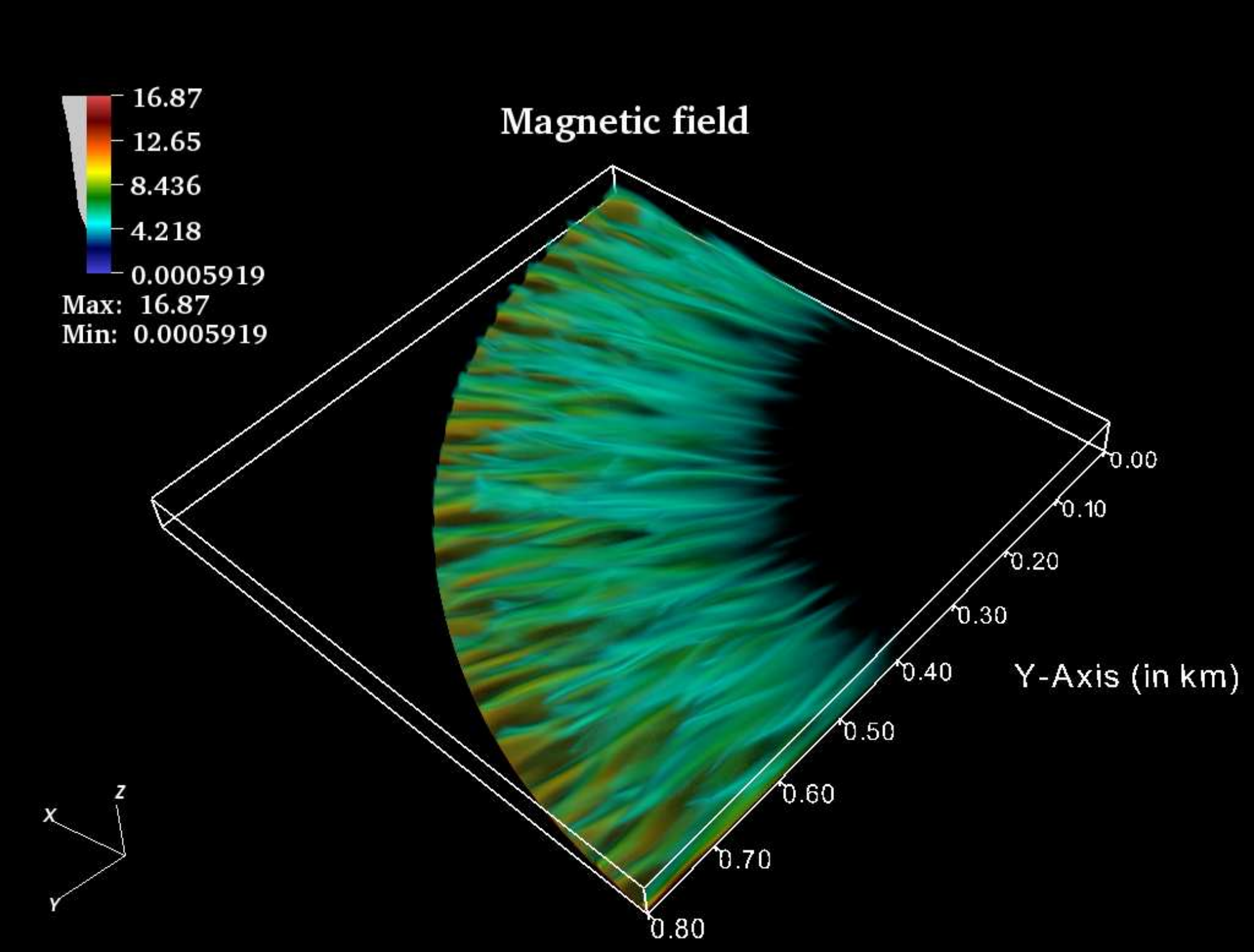}
	\includegraphics[width = 8.5cm, height = 6.60875cm] {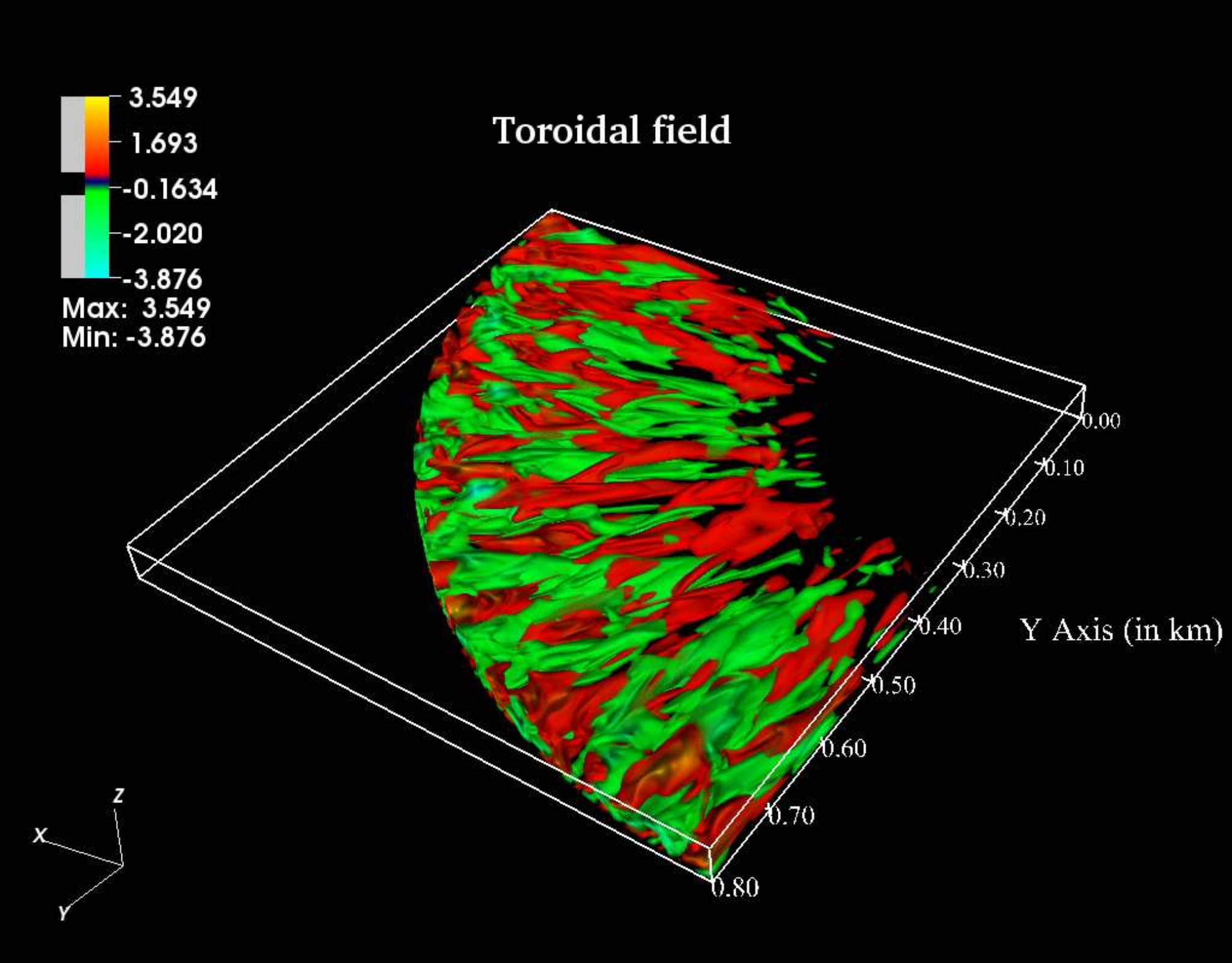}
	\caption{\small Left: 3D contours of magnetic field magnitude at $t\sim 2t_A$ for $Z_c=70$m filled mound with $B_p=10^{12}$G. The field values are in units of $10^{12}$G. Alternate channels are clearly seen to form magnetic valleys which coincide with density streams in Fig.~\ref{density70m}. Right: 3D contours of $B_\phi$ showing alternate strips of positive and negative toroidal components signifying turbulent eddies.}\label{field70m}
\end{figure*}
The final form of the density distribution is shown in Fig.~\ref{density70m}. Finger-like streams are clearly seen to extend throughout the mound. The magnetic field also develops finger-like channels across the azimuthal domain with alternate regions of high and low field, which are complementary to the density streams (see Fig.~\ref{field70m}). Structures of higher density settle in valleys between the magnetic channels as matter tries to flow out radially. Although poloidal field is dominant, there is a substantial build up of toroidal fields from an initial zero value.  As seen in Fig.~\ref{field70m}, the toroidal field is intermittent with alternate strips of negative and positive regions signifying local eddies and churning of internal fields.

\begin{figure}
	\centering
        \includegraphics[width = 7.cm, height = 7.cm,keepaspectratio] {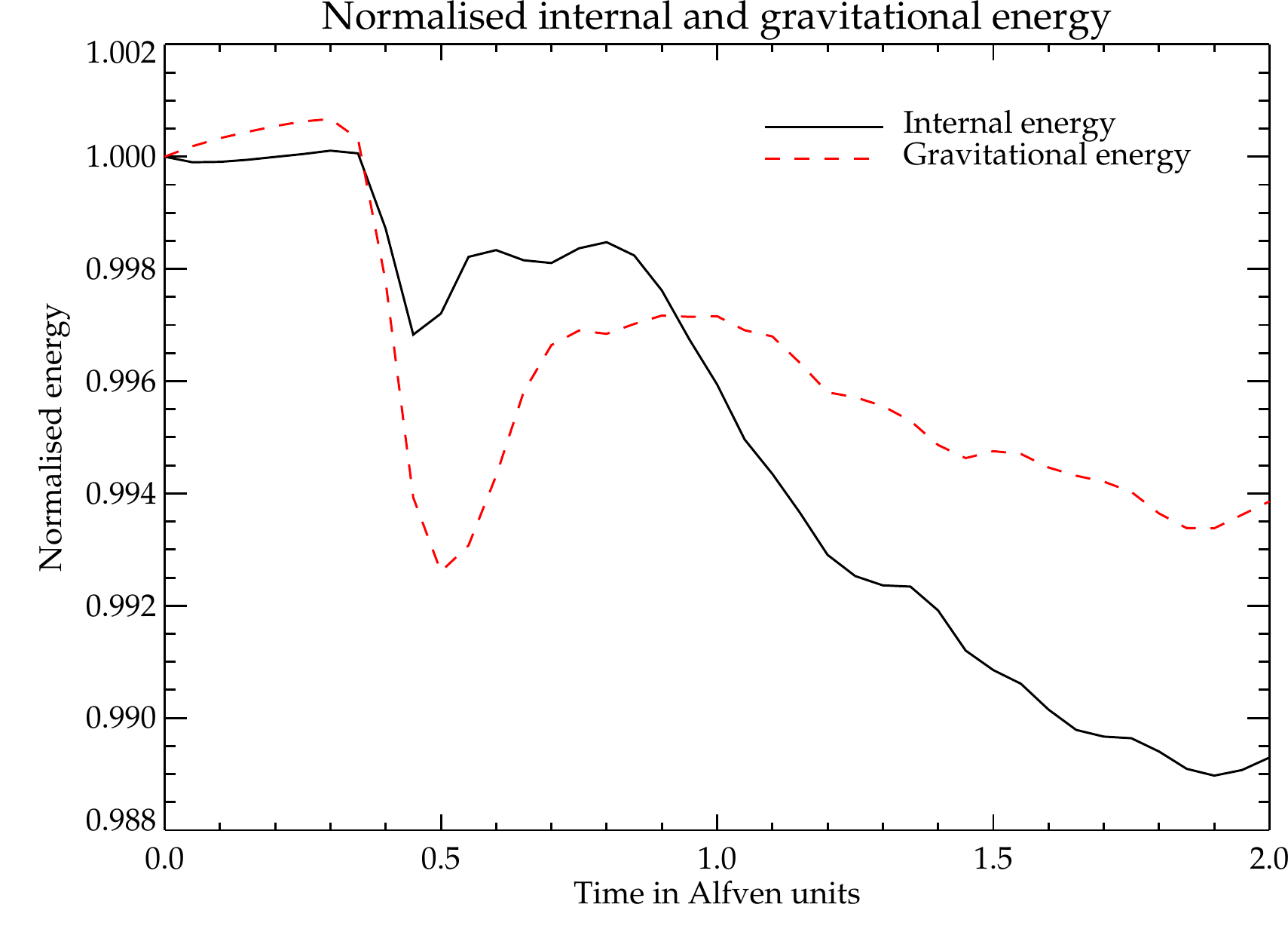}
	\caption{\small Internal and gravitational potential energy normalised to their initial values $\sim 2.76\times 10^{23}$ ergs and $\sim 3.21\times 10^{23}$ergs respectively. } \label{cumulen70m}
\end{figure}
{\bf Energy:} The total energy of the mound is dominated by its internal and gravitational potential energy components, magnetic energy contributing only a small fraction ($\sim 5.5 \%$ of internal energy at $t=0 $). The various energy components remain almost constant with very little change till the instability onset time\footnote{Defined here as the time at which the magnitude of the rate of change of the energy components increases by more than a factor of ten of the initial value } $t_{\rm inst}\sim 0.3 t_A$ (see Fig~\ref{cumulen70m}),  beyond which the rate of change of the energy components increases due to the onset of MHD instabilities. This also corresponds to the time when the finger-like radial channels are formed throughout the mound. Thus for the $\sim 70$m mound, the instabilities start acting after $ \sim 0.3 t_A$. The internal and gravitational energy decrease as some part is converted to magnetic energy and the rest is lost through outflows from the periphery. The system is dominated by poloidal field components. Although toroidal fields are generated due to internal motions, they contribute only a maximum of $2.3\%$ of the total magnetic energy. As the instabilities saturate, stretching and twisting of field lines increases the magnetic energy ($\sim 20\%$ increase at $t\sim 2 t_A$). 

\begin{figure}
	\centering
        \includegraphics[width = 7.cm, height = 7.cm,keepaspectratio] {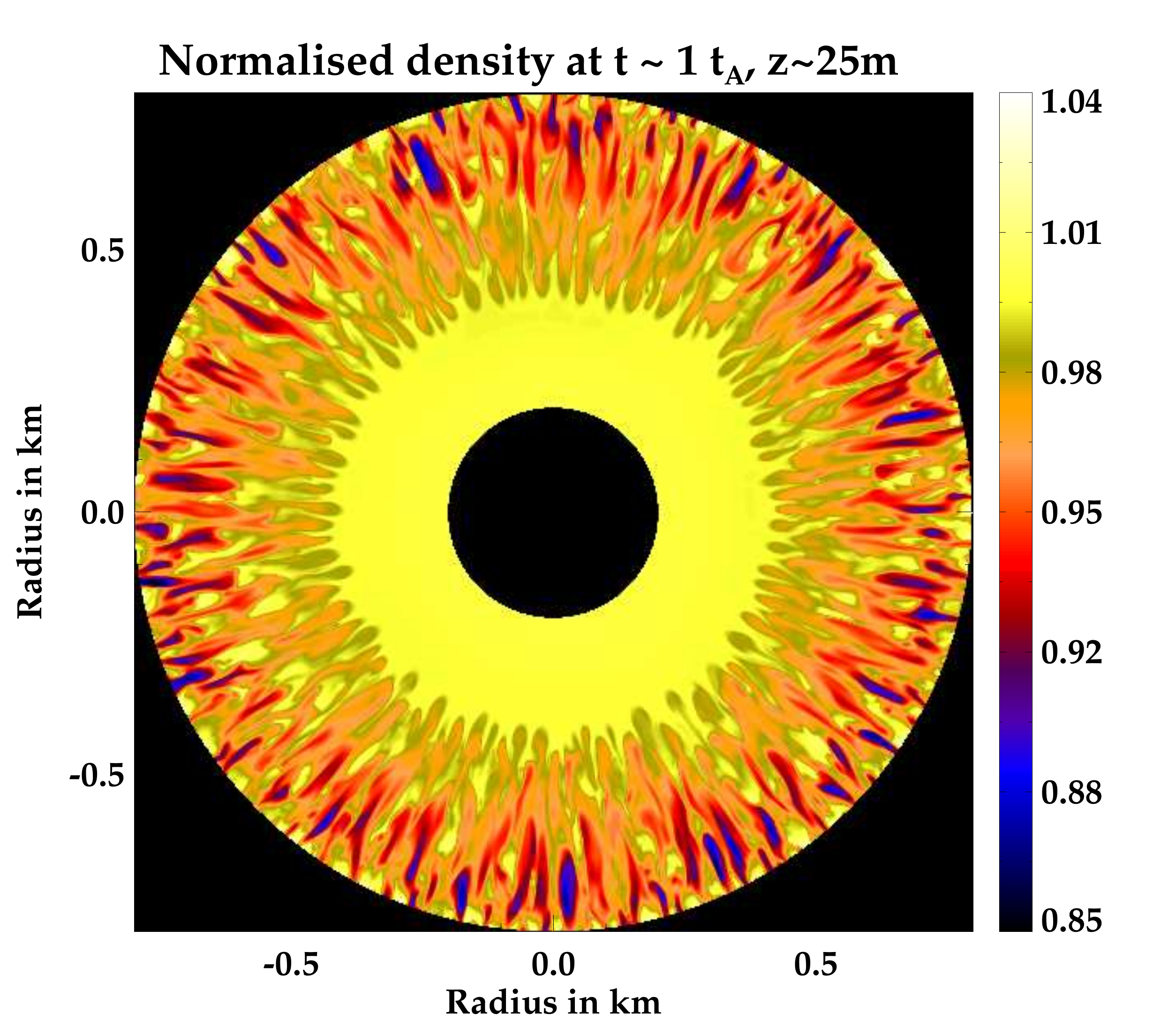}
	\caption{\small Density of a 70m mound at $z\sim 25$m, normalised to equilibrium value, for a simulation run with azimuthal domain: $[0,2\pi]$. The entire mound develops alternating density channels in the azimuthal direction destroying the axisymmetry of the equilibrium solution.} \label{fullrho}
\end{figure}
\begin{figure}
	\centering
        \includegraphics[width = 7.cm, height = 7.cm,keepaspectratio] {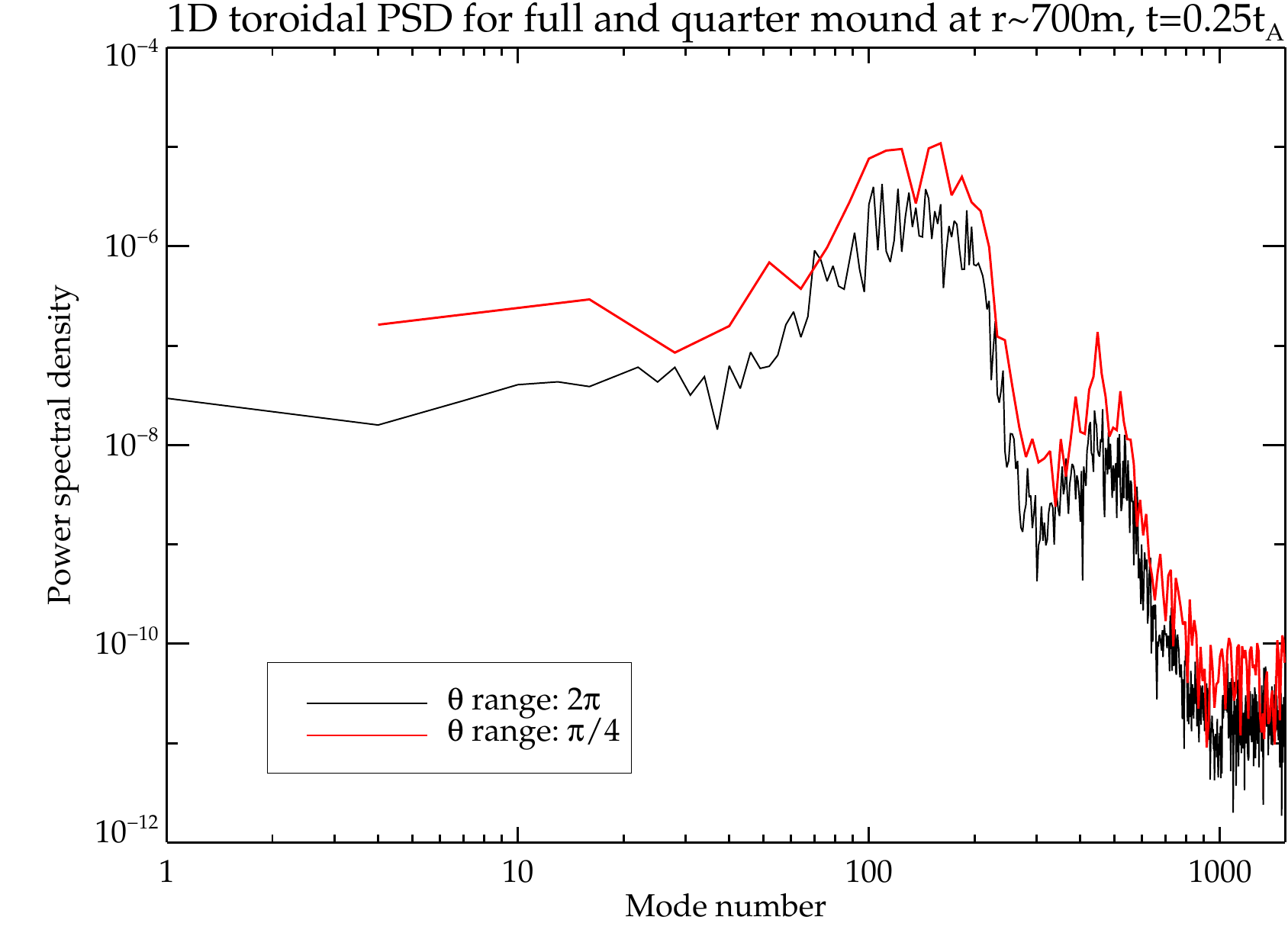}
	\caption{\small Power spectral density of 1D FFT of perturbed density at $r\sim 700$m, $z\sim 25$m and $t\sim 0.25 t_A$ for simulation runs with azimuthal domains $[0,2\pi]$ (in black) and $[0,\pi/2]$ (in red). The PSD of the two runs closely resemble each other confirming that growth of MHD instabilities are qualitatively similar in both cases.} \label{PSDcompare}
\end{figure}

{\bf $\boldsymbol{2\pi}$ vs $\boldsymbol{\pi/2}$ in azimuthal domain:} As the runs were computationally expensive, we have performed most of the simulation over a quadrant of the azimuthal domain. This assumes periodic boundary conditions at $\theta = 0$ and $\theta = \pi /2$. Such an approximation does restrict some large $m$ modes from growing. However if the dynamics is dominated by local variations in physical parameters, then short wavelength modes are expected to dominate. To test the validity of our results, we have performed a few simulations for the full azimuthal domain $[0,2\pi]$ for a smaller evolution time and compared them with single quadrant simulations. The results are qualitatively similar in both cases with similar nature of the growth of the instabilities (see Fig.~\ref{fullrho} for normalised density at $\sim 1t_A$).  It is seen that the PSD of the physical parameters (e.g. density) at different heights and radii are very similar for both cases (see Fig.~\ref{PSDcompare} for a comparison at $t \sim 0.25 t_A$), indicating that the results of our simulations are not significantly affected by constraining the domain to $[0,\pi/2]$. Hence further runs have been performed over the $[0,\pi/2]$ domain of the azimuthal angle.

\begin{figure*}
	\centering
        \includegraphics[width = 5.8cm, height = 5.5cm] {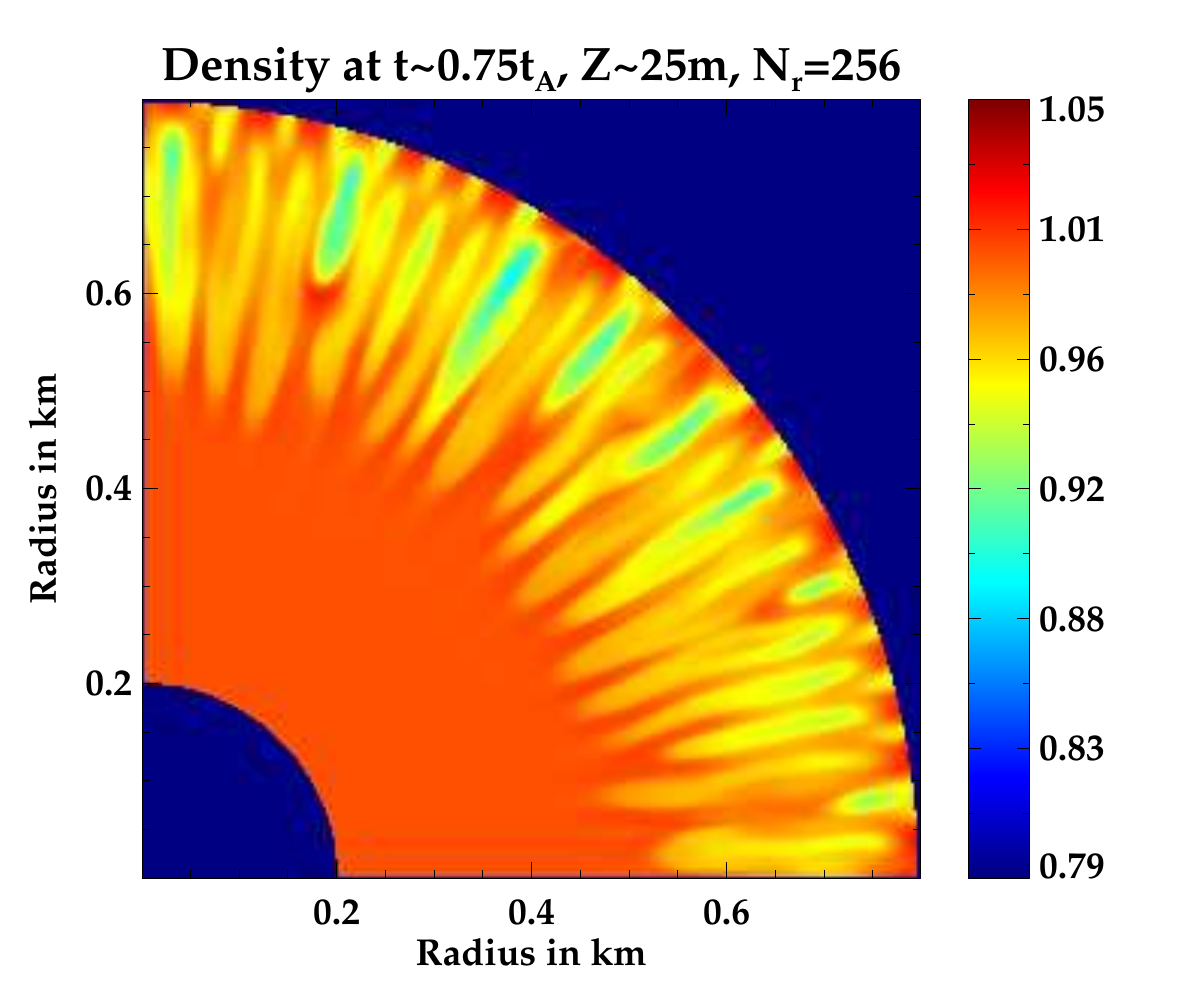}
	\includegraphics[width = 5.8cm, height = 5.5cm] {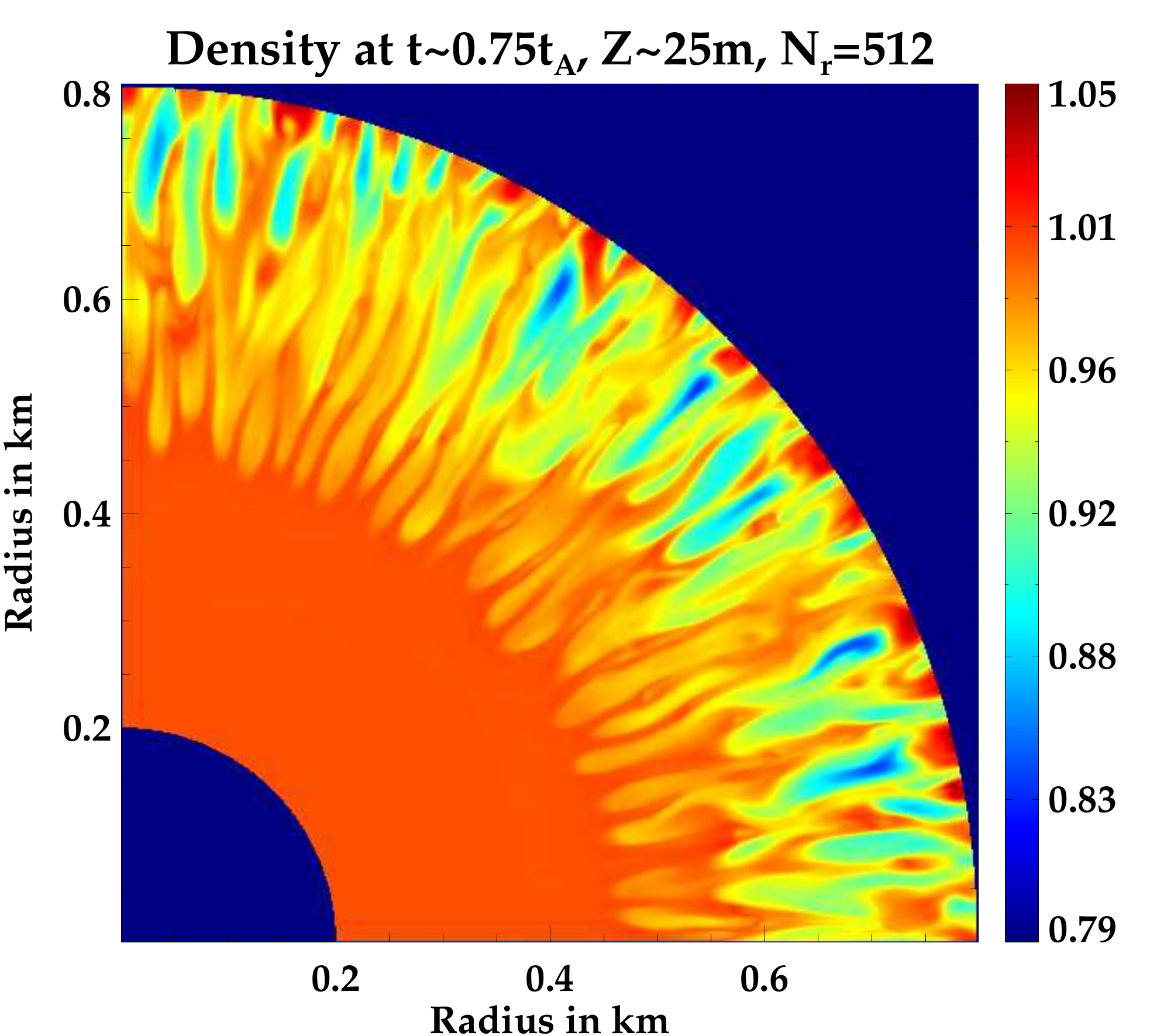}
	\includegraphics[width = 5.8cm, height = 5.5cm] {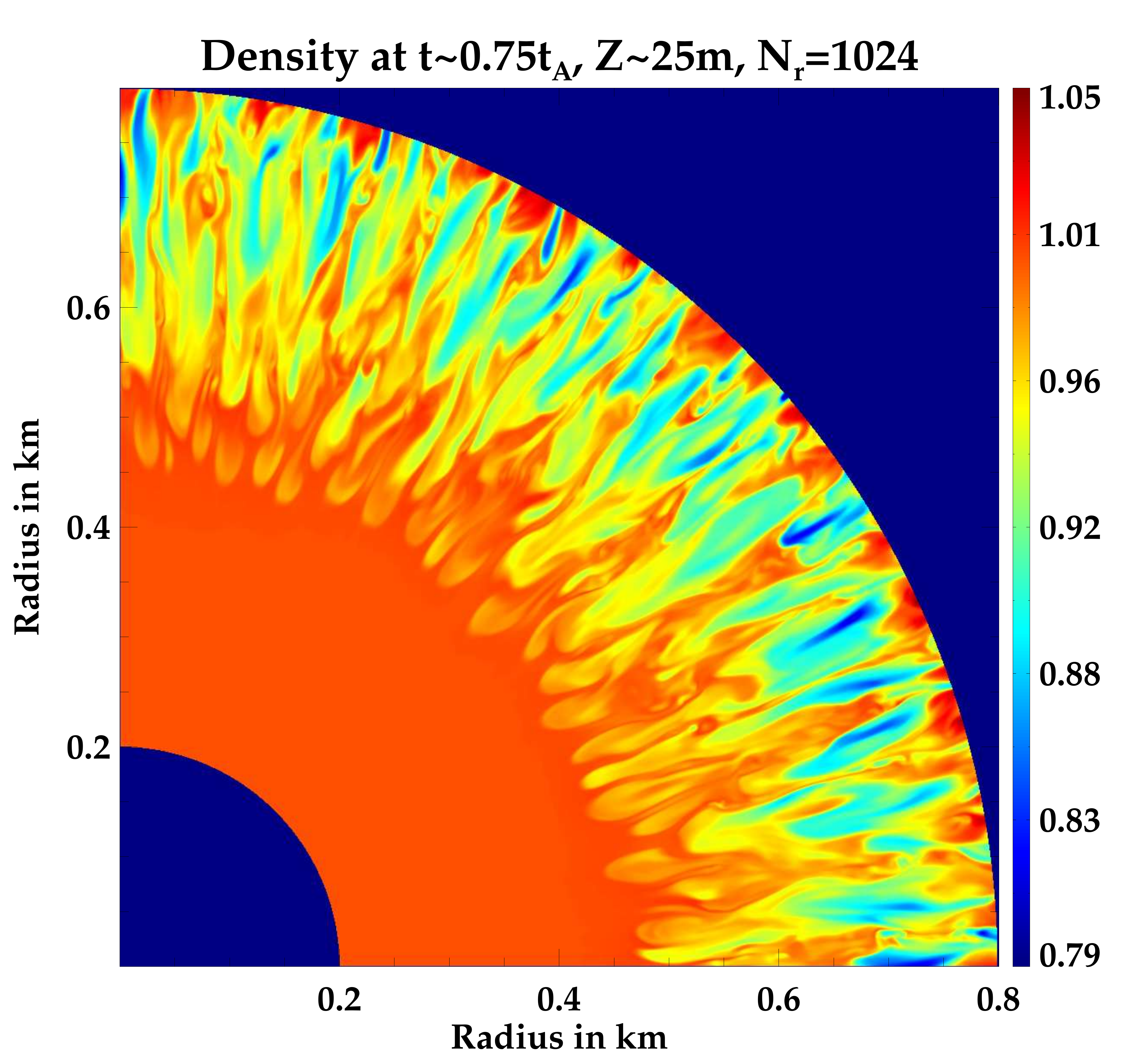}
	\caption{\small Density normalised to equilibrium value at $t\sim 0.75 t_A$ and $z\sim 25$m for different resolutions ($N_r\times N_\theta \times N_z$): $256\times 384 \times 24$, $512\times 768\times 48$ and $1024 \times 1536 \times 96$ respectively.}\label{rhorescompare}
\end{figure*}
\begin{figure}
	\centering
        \includegraphics[width = 7.cm, height = 7.cm,keepaspectratio] {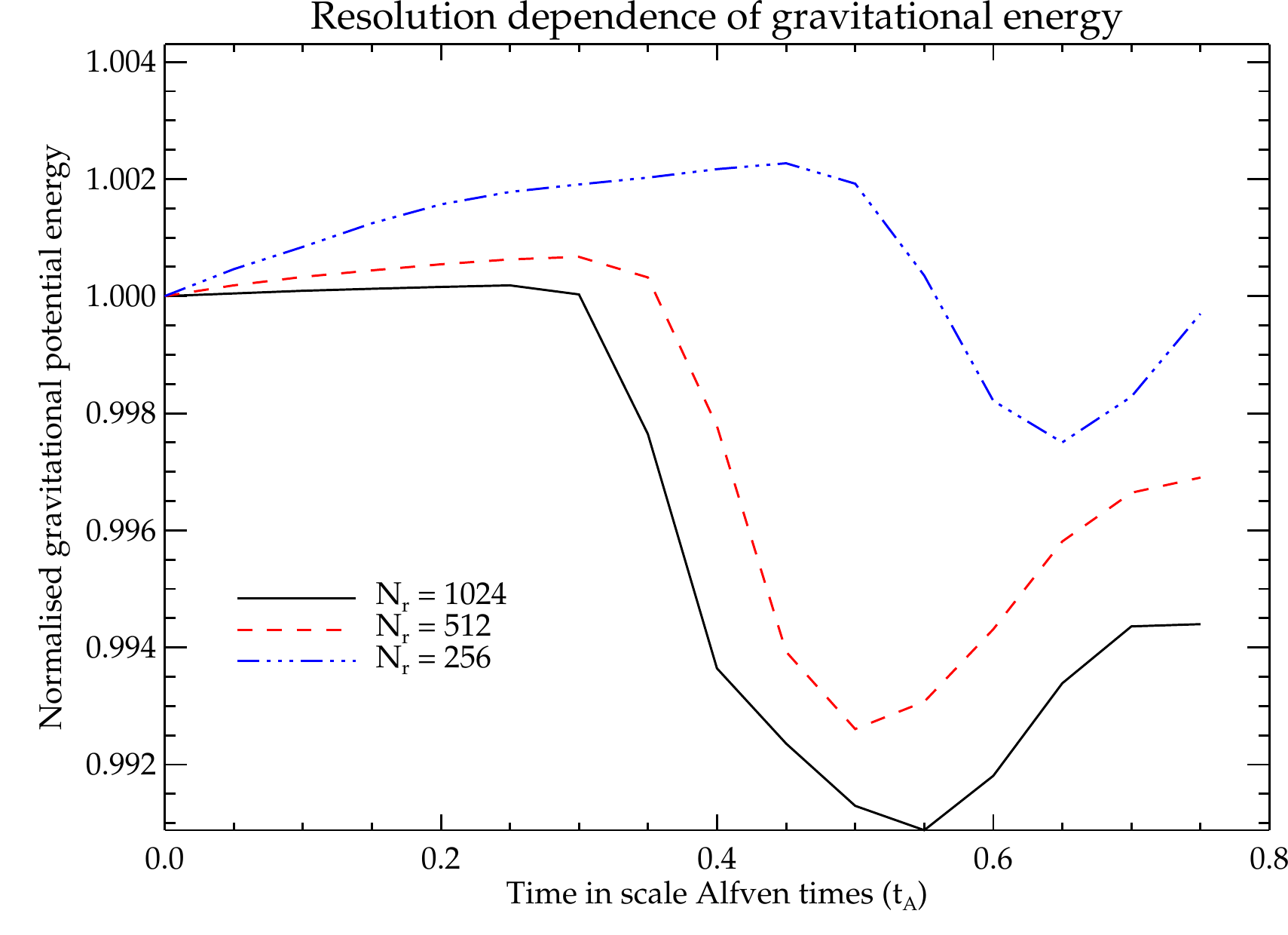}
	\caption{\small  Evolution of gravitational potential energy for three different resolutions (as in Fig.~\ref{rhorescompare}) with number of radial cells as: $N_r=256$, $N_r=512$ and $N_r=1024$. See text for details.} \label{gravcompare}
\end{figure}

{\bf Resolution dependence:} We have performed the simulation at different resolutions to check for convergence. At higher resolutions more internal structures are resolved. As cross field diffusion is reduced, field lines are deformed at earlier times due to internal flows\footnote{As we work in the ideal MHD limit, dissipation is numerical in nature, and decreases with increase in resolution. In real systems this will be controlled by resistive effects at small scale.}. Thus higher resolutions result in faster onset of the instabilities, which has also been demonstrated for 2D simulations in MBM13. However, we have found that for grids with spatial resolutions $\leq 1$m, the results of the MHD instabilities are qualitatively similar. In Fig.~\ref{rhorescompare} the density at $t\sim 0.75 t_A$ and $z\sim 25$m is plotted for three different resolutions ($N_r\times N_\theta \times N_z$): $256\times 384 \times 24$, $512\times 768\times 48$ and $1024 \times 1536 \times 96$ respectively. It is seen that for the simulation with radial grid size $N_r=256$ the radial structures have not developed yet whereas for the simulations with $N_r=512$ and $N_r=1024$ the toroidal modes have fully developed and reached the non-linear saturation phase. The time evolution of the gravitational potential energy at different resolutions are compared in Fig.~\ref{gravcompare} (other energy components behave similarly). There is a tendency for the energy curves to converge at higher resolution. It can be seen that the runs with $N_r=512$ and $N_r=1024$ resemble each other better than that with $N_r=256$. Thus further simulations were performed with the resolution set to $512\times 768 \times 48$, which corresponds to a spatial resolution of $\sim 1$m at the centre of the domain. Similar resolution settings have been adopted for mounds of different sizes.

\subsection{Mounds of medium size : threshold of stability}
\begin{figure}
	\centering
        \includegraphics[width = 7.cm, height = 7.cm,keepaspectratio] {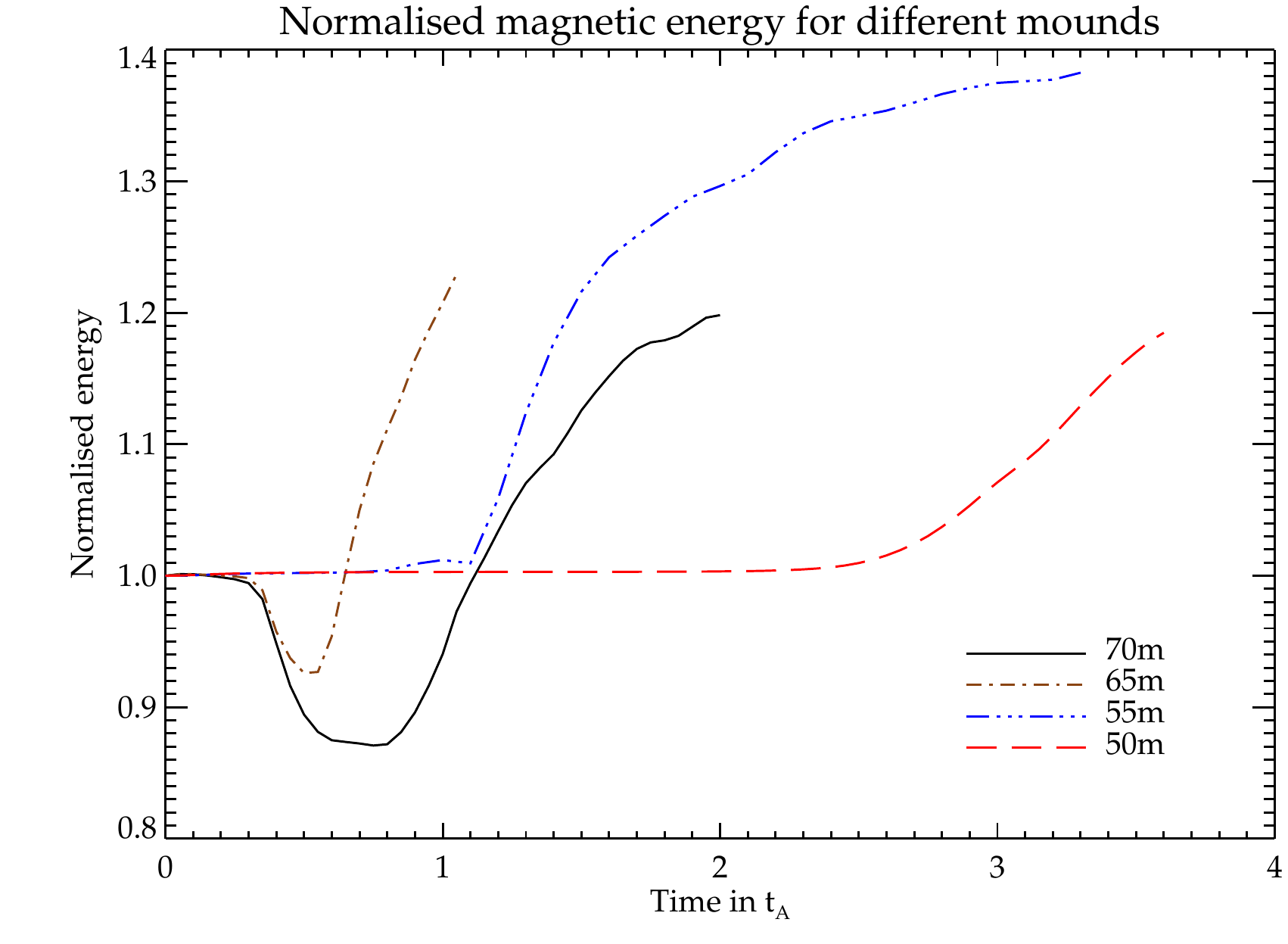}
	\caption{\small Magnetic field energy for different mound heights. The runs have been evolved till the MHD instabilities have sufficiently developed. The instability onset time $t_{\rm inst}$ (see text) is greater for larger mounds. } \label{multimounden}
\end{figure}
\begin{figure*}
	\centering
        \includegraphics[width = 5.8cm, height = 5.5cm] {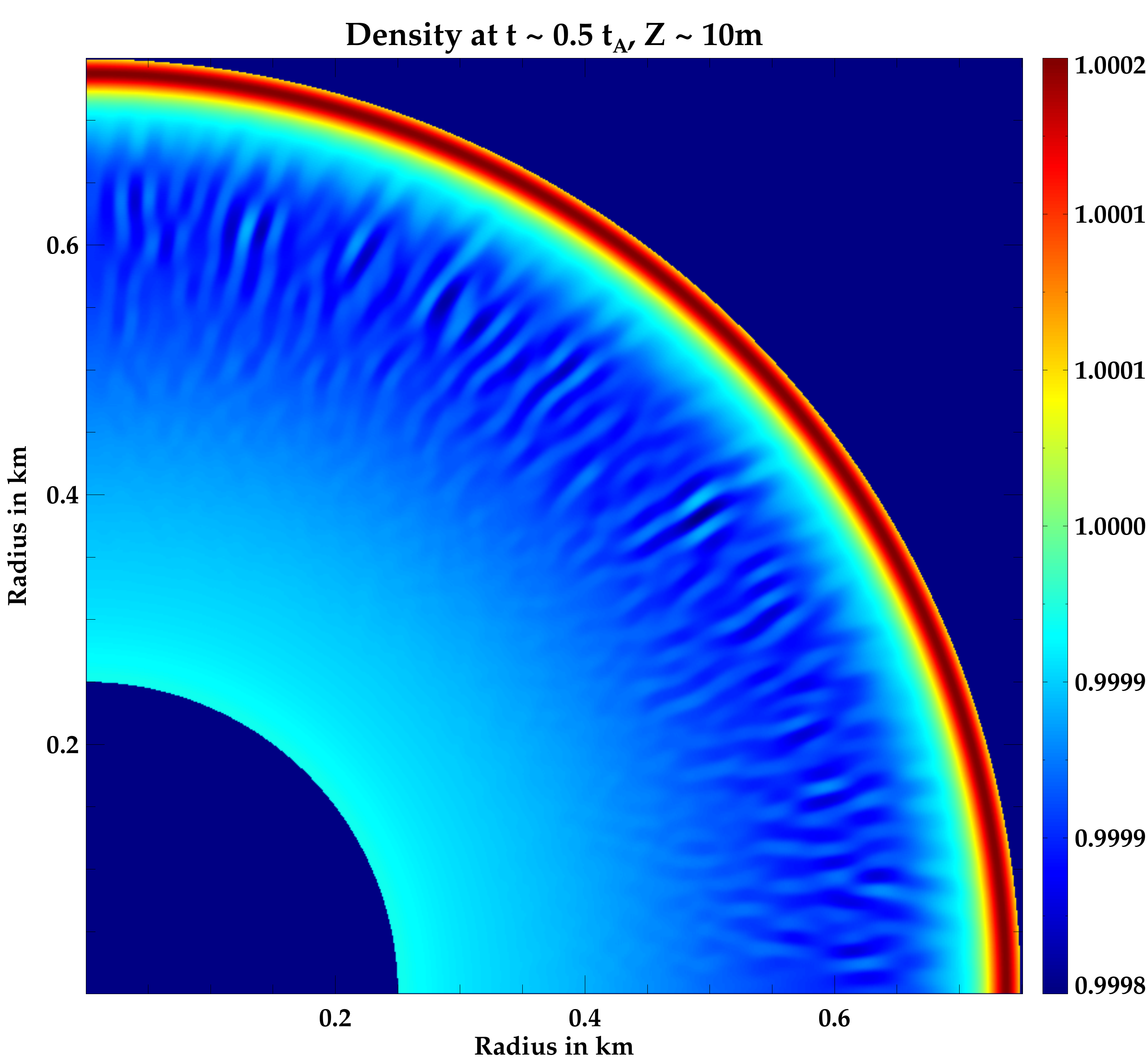}
	\includegraphics[width = 5.8cm, height = 5.5cm] {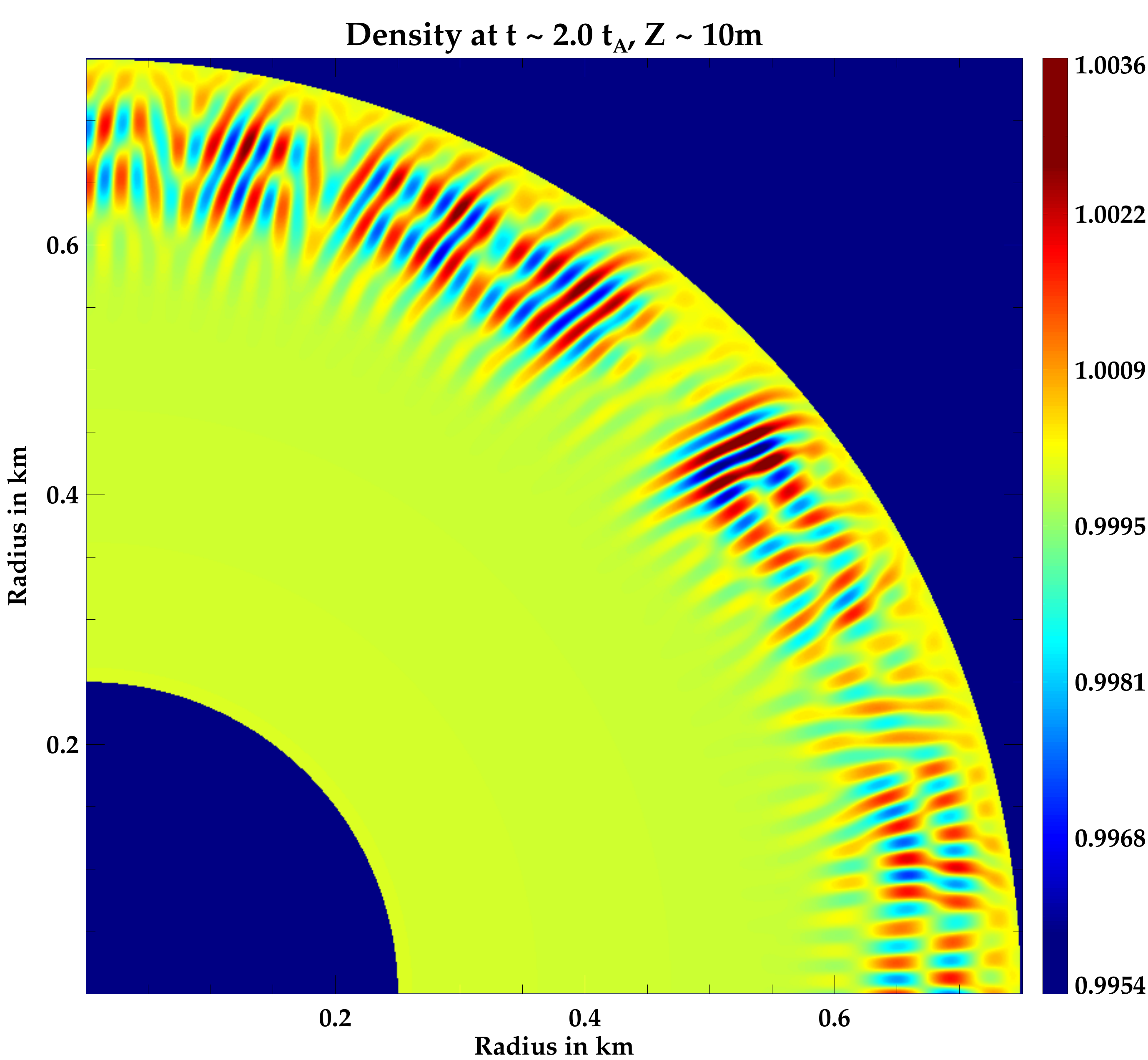}
	\includegraphics[width = 5.8cm, height = 5.5cm] {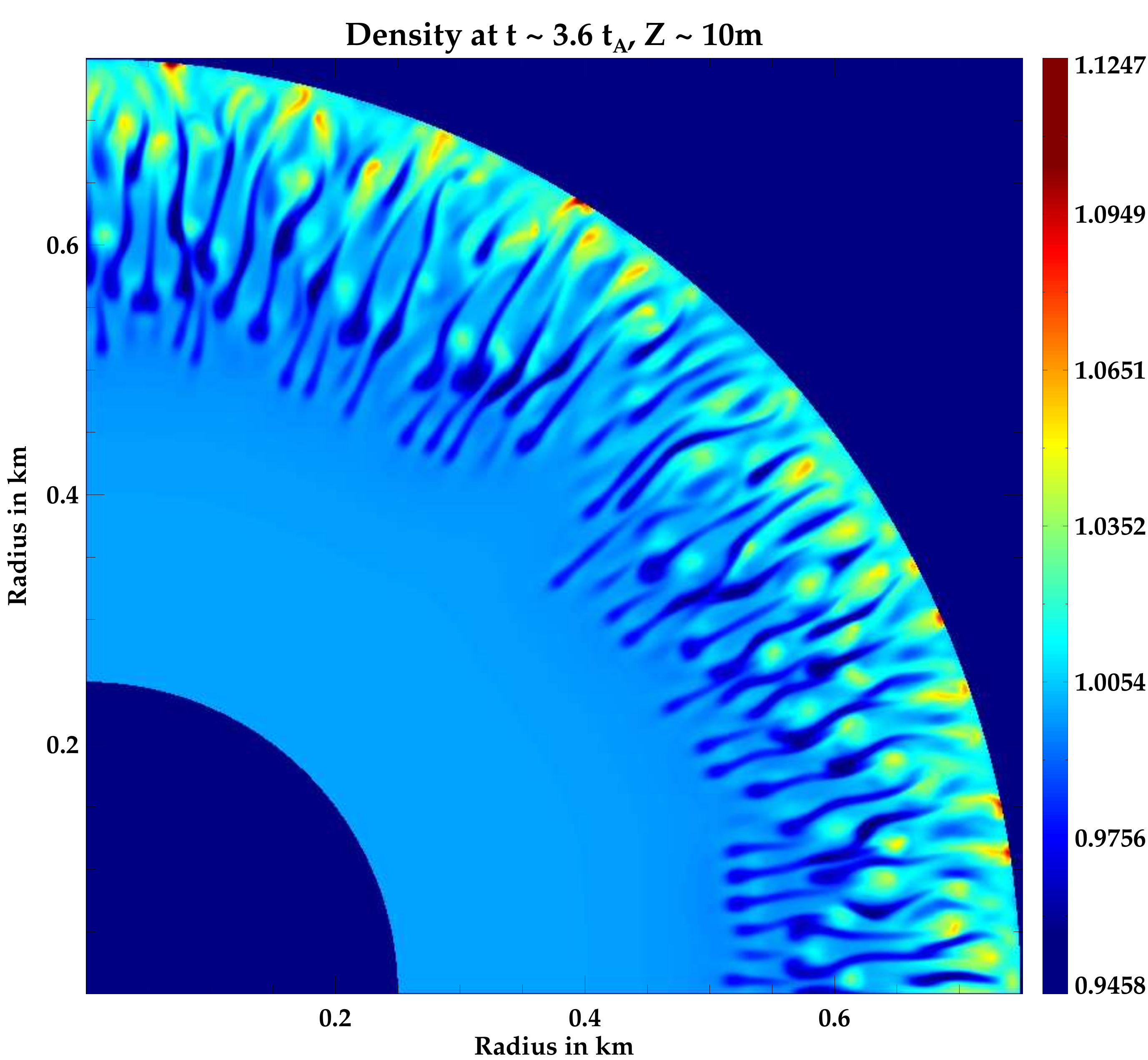}
	\includegraphics[width = 5.8cm, height = 5.5cm] {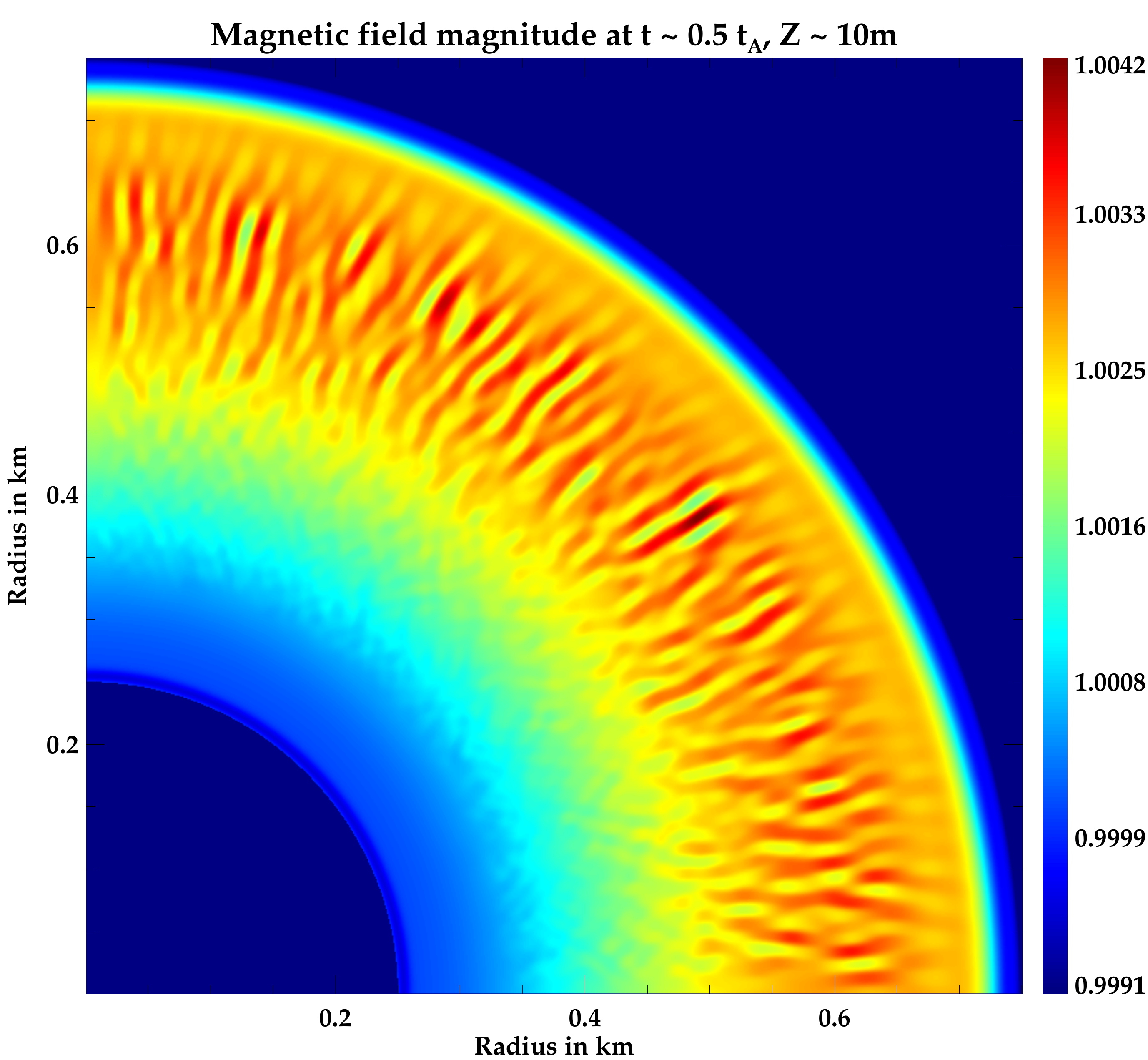}
	\includegraphics[width = 5.8cm, height = 5.5cm] {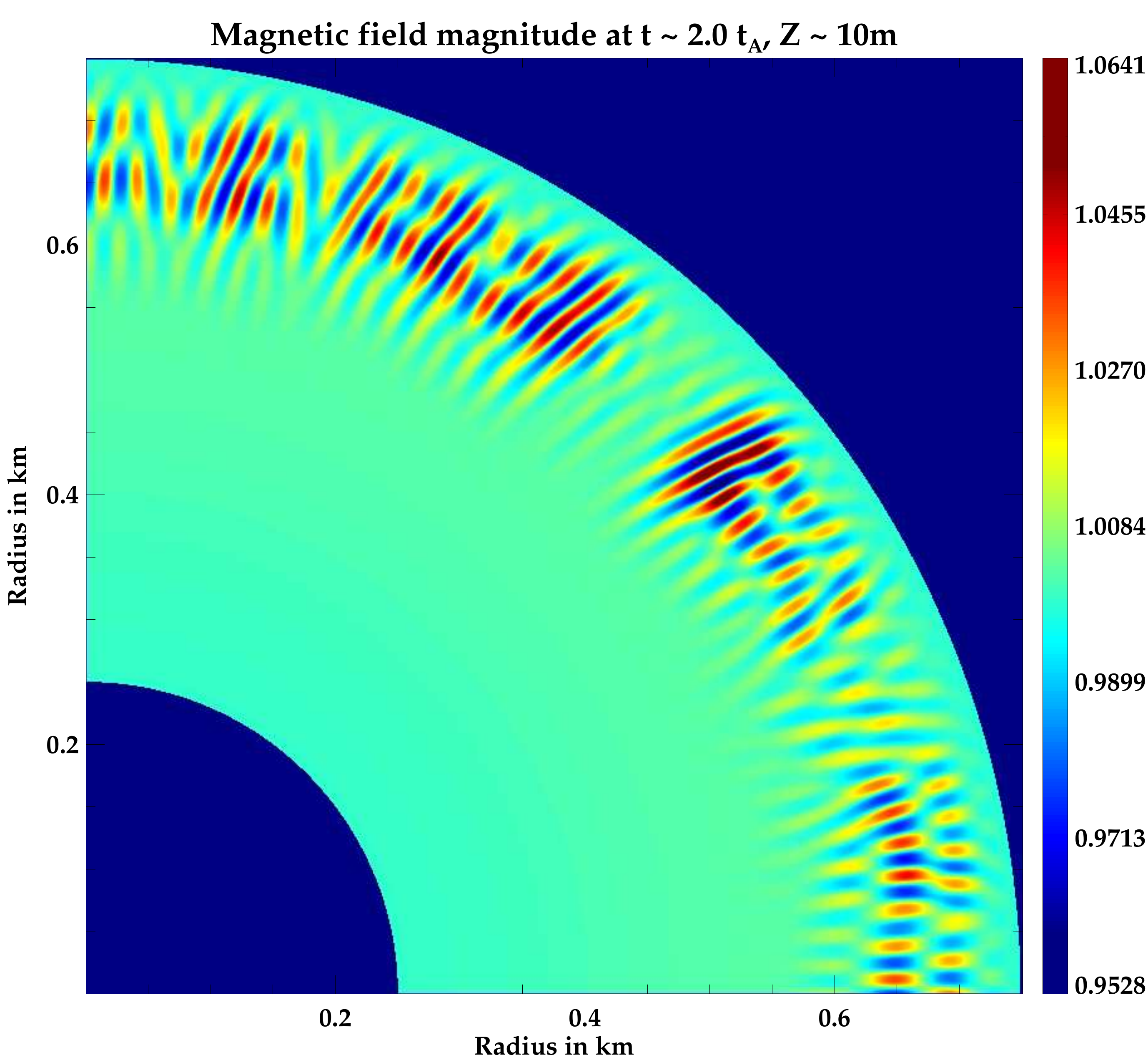}
	\includegraphics[width = 5.8cm, height = 5.5cm] {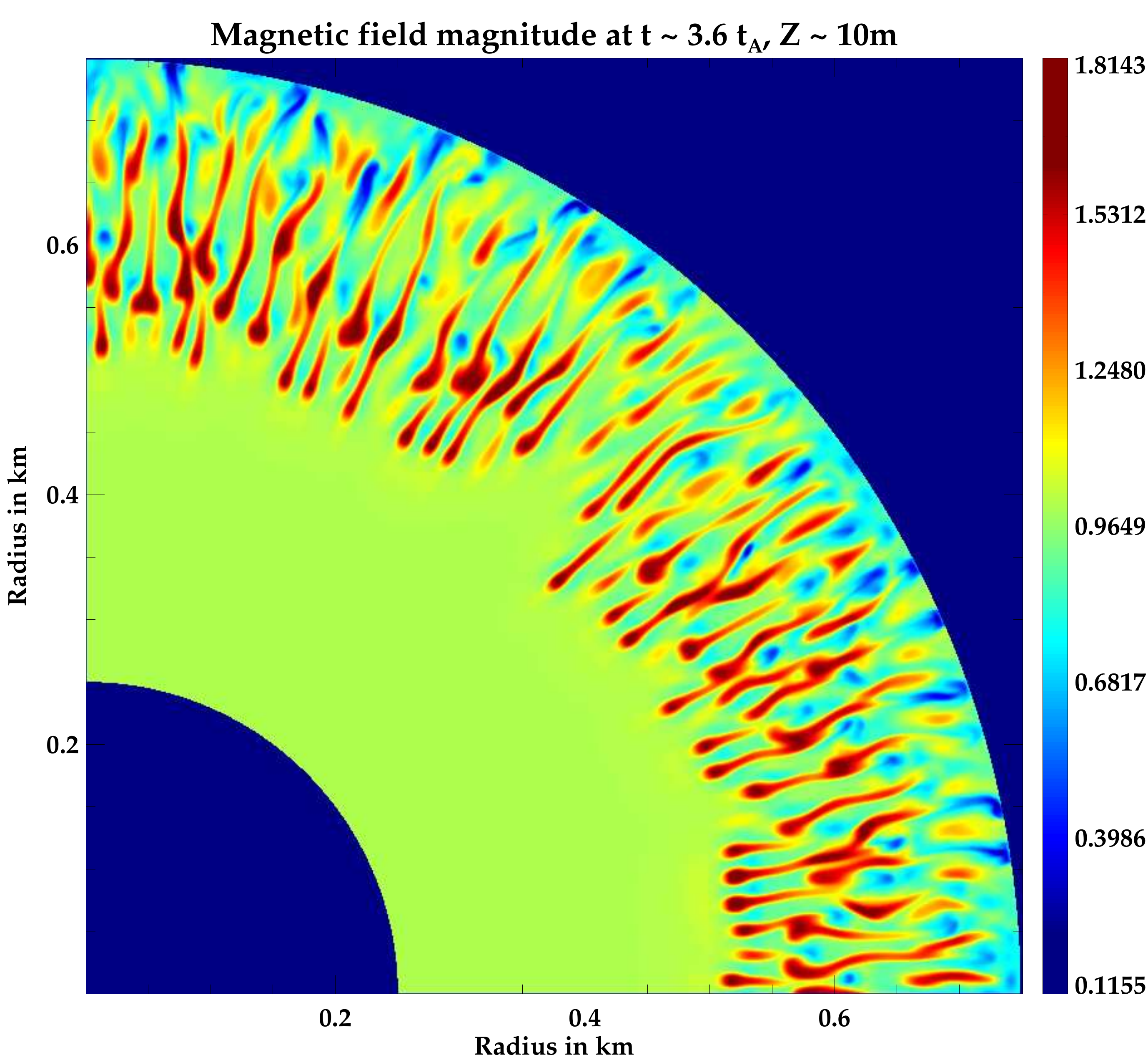}
	\caption{\small Top: Density normalised to equilibrium value at $z\sim10$m at different times. Bottom: Magnetic field magnitude normalised to equilibrium value at $z\sim10$m. At the initial phases the density settles in stationary pockets of low field regions, which starts to spread at later times.  }\label{50mrhobmag}
\end{figure*}
\begin{figure}
	\centering
        \includegraphics[width = 7.cm, height = 7.cm,keepaspectratio] {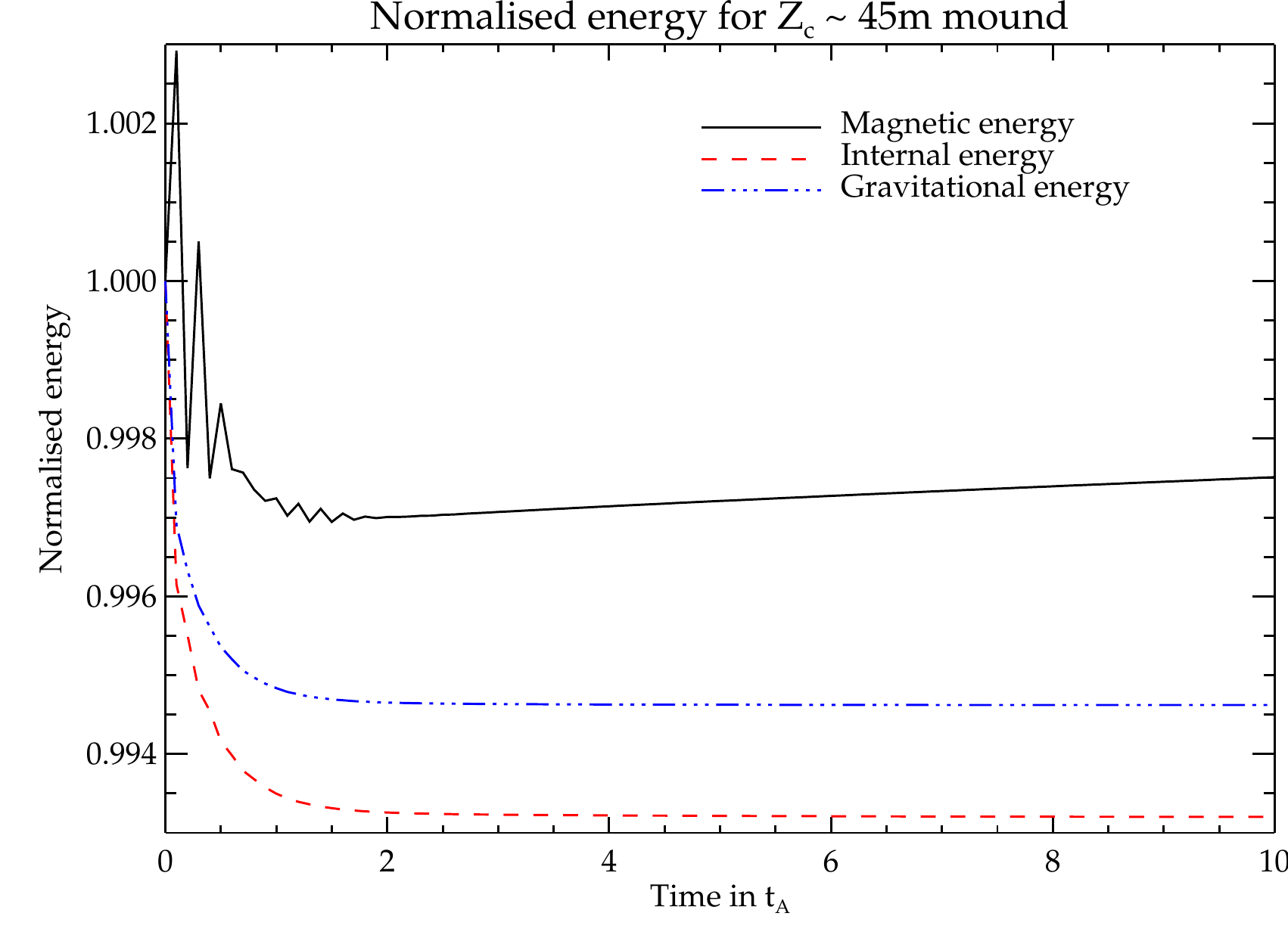}
	\caption{\small The energy components for $Z_c=45$m mound normalised to their initial values: internal energy to $\sim 5.1\times 10^{23}$ergs, gravitational potential energy to $\sim 3.3\times10^{23}$ergs and magnetic energy to $\sim 8.8\times10^{21}$ergs. The internal and gravitational energy relaxes to a stationary state after a very small change from initial value. The magnetic energy also settles down after some initial oscillations. The magnetic energy in the steady state is seen to have a slight increasing trend over $\sim 10 t_A$, but the change is very small ($\sim 0.05 \%$). }\label{45mben}
\end{figure}
Mounds of smaller mass with less field distortion display slower growth of instabilities. For mounds near the GS threshold (e.g. $Z_c\sim 70{\rm m}$ and $ 65{\rm m}$) the instability onset time is $t_{\rm inst}\sim 0.3 t_A$ and $\sim 3.5 t_A$ respectively, whereas for smaller mounds it is larger e.g. $t_{\rm inst}\sim 1.1 t_A$ for $Z_c=55$m and $\sim 2.5 t_A$ for $Z_c=50$m (see Fig.~\ref{multimounden}).  For a 50m mound, the density initially settles in stationary pockets of lower field strength (see Fig.~\ref{50mrhobmag}). The initially formed filaments remain stationary till $\sim 2.5 t_A$ after which the instability starts to spread throughout the mound with a corresponding change in the energy components (see Fig.~\ref{multimounden} for magnetic energy plot).

For a $Z_c=45$m mound, MHD instabilities do not grow during the run time of the simulation. The perturbed mound quickly settles into a steady state after some initial oscillations (see Fig.~\ref{45mben}). Thus with decrease in mound mass, the system tends to become stable. This agrees with the results of \citet{litwin01} which predicts that accretion mounds are unstable only if plasma $\beta = p/(B^2/8\pi)$ is higher than a threshold value. As shown in the discussion of MBM13, the maximum plasma $\beta$ for a $Z_c=45$m mound is very close to such a threshold. This confirms that filled mounds of central height lower than $\sim 45$m may be stable with respect to pressure driven ballooning instabilities.

\subsection{Hollow mound}
\begin{figure}
	\centering
        \includegraphics[width = 7.cm, height = 7.cm,keepaspectratio] {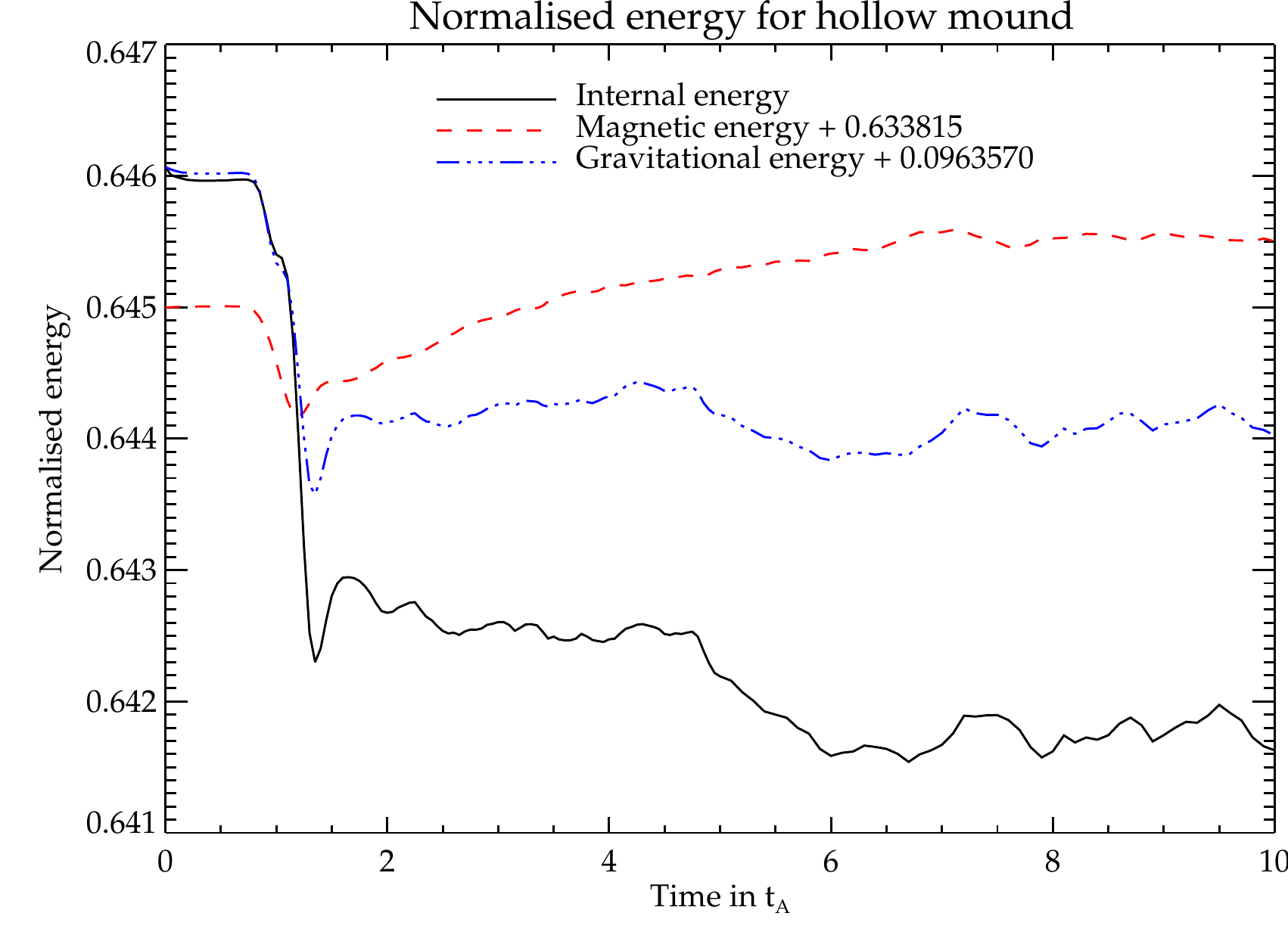}
	\caption{\small Energy components of a simulation run with hollow mound normalised to $E_0=7.95\times10^{22}$ ergs. The magnetic energy is plotted with an offset of $\sim 0.096E_0$ and gravitational potential energy with an offset of $\sim 0.63 E_0$ to represent all three components in the same scale. The magnetic energy is seen to increase by $\sim 5\%$ before reaching a plateau in the non-linear saturation phase.} \label{holoen}
\end{figure}

\begin{figure*}
	\centering
        \includegraphics[width = 8cm, height = 8cm,keepaspectratio] {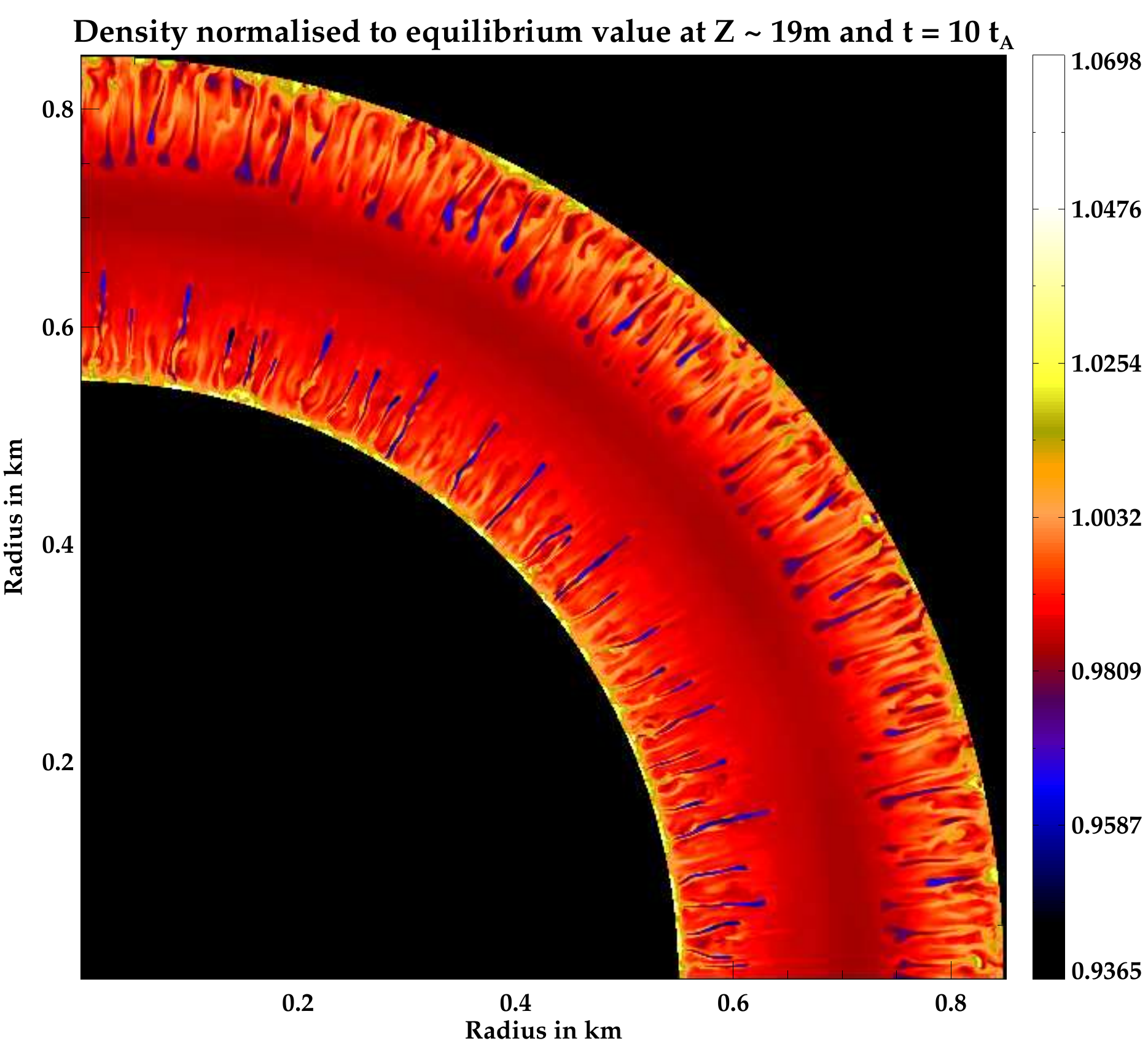}
        \includegraphics[width = 8cm, height = 8cm,keepaspectratio] {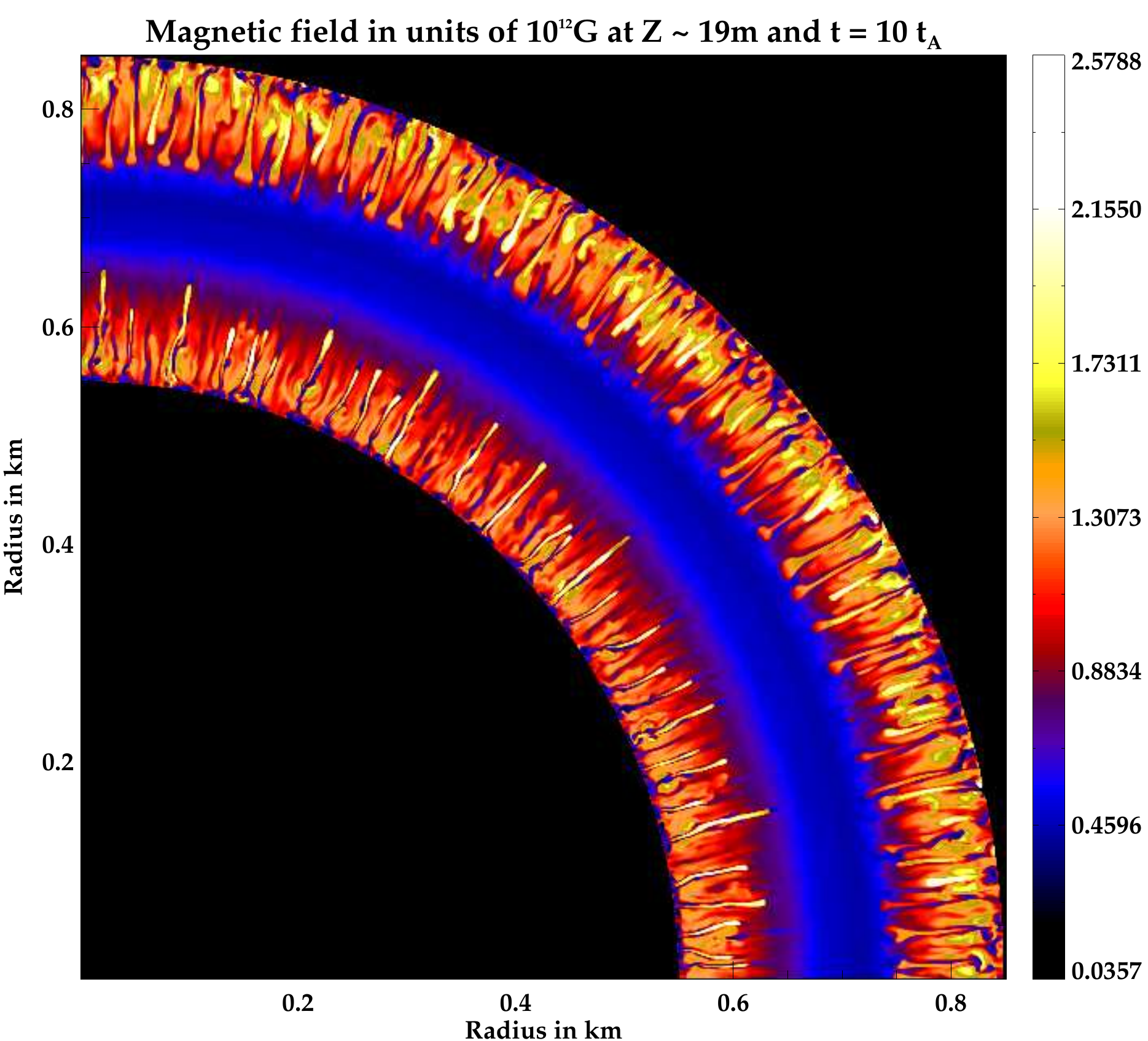}
	\caption{\small Left: Density (normalised to equilibrium value) of a hollow mound at $t\sim 10 t_A$ and $z\sim 20$m. The accretion geometry modelled is that of a ring-like mound with hollowed interior. Right: Corresponding magnetic field magnitude in units of $10^{12}$G. The finger like streams throughout the domain shows the spread of MHD instabilities at both outer and inner radii of the hollow mound. } \label{holoslices}
\end{figure*}

As discussed in MBM13, a finite range of the mass loading region in the accretion disc will result in the formation of hollow ring-like accretion mounds on the neutron star surface. The GS solutions and results from 2D perturbation tests have already been discussed in MBM13. Here we perform 3D perturbation tests of GS solutions of such mounds (eq.~(\ref{holopro}) with $z_c = 45$m). As in filled mounds, multiple radial streams appear at the outer edge. The instabilities saturate by about $\sim 2 t_A$, as is evident from the flattening of the energy curves after initial changes (see Fig.~\ref{holoen}). However, unlike in filled mounds, radial streams also appear at the inner boundary (see Fig.~\ref{holoslices}), which is characterised by a second dip in the internal energy at  $\sim 5 t_A$. The toroidal mode instabilities will cause radial outflow of matter at both outer and inner radial edges, which may, in time, fill up the interior of the hollow mound. However, due to constraints of compute resources, we have not been able to evolve the system for long enough to detect significant mass loss in our current runs. In any case, the restrictive boundary conditions used in our simulations would be unable to provide an adequate description of mass outflow.

\section{Discussion}\label{Sec.discuss}
In this paper we have performed 3D MHD simulations of perturbed accretion mounds which are initially in force equilibrium. We have confirmed the onset of pressure driven toroidal mode instabilities. \citet{litwin01} had shown that $m\neq0$ non-axisymmetric modes will disrupt the equilibria for mounds beyond a threshold mass. Such ballooning type instabilities are multi-dimensional in nature and require a full 3D simulation for their investigation. Our simulations show that such instabilities result in the formation of multiple radially elongated streams distributed in the azimuthal direction. Part of the internal and gravitational potential energy is converted to magnetic energy and the strength is increased by the stretching of field lines due to internal motions. There is a substantial increase of the toroidal magnetic field component (from initial zero value). The instabilities grow more slowly in mounds of smaller size, and for mounds smaller than a threshold, instabilities were not excited within the run time of the simulations. This roughly corresponds to the threshold discussed by \citet{litwin01} e.g. for a $Z_c=45$m mound, from our GS solutions we have $\beta _{\rm max}=293$, which is close to the threshold $\beta$ ($\sim 260$) predicted by \citet{litwin01} for MHD instabilities (see discussion in MBM13).

A threshold mound mass was obtained by MB12 from static solutions, which was shown in MBM13 to correspond to the mass threshold beyond which gravity driven instabilities are triggered. From our current 3D MHD simulations we predict a lower mass threshold ($\sim 5\times10^{-13} M_\odot$ for $B_p \sim 10^{12}$G) above which pressure driven instabilities will start operating and matter will not be efficiently confined by the local field in the polar cap. Further addition of mass will result in the formation of radial density streams nestled in magnetic valleys and matter will flow out of the polar cap to spread over the surface of the neutron star. Our current simulations, limited by computational resources, could not follow the evolution of the mound beyond a few Alfv\'en times. However, we can expect that with continued accretion, there will be a dynamic equilibrium between inflow from the top and outflow from the sides. Such an outflow, aided by MHD instabilities, may prevent the secular process of field burial often invoked to explain the low magnetic fields of neutron stars in LMXBs and millisecond pulsars.

Semi-analytic work and MHD simulations by \citet{melatos04} (hereafter PM04), \citet{payne07} and \citet{vigelius08} address the problem of field burial by accretion and come to the conclusion that mounds of size $\sim 10^{-4} M_\odot$ can cause the field to be buried. However our current work differs from the approach of PM04 in various aspects. We consider the confined matter to be a degenerate Fermi plasma whose pressure is several orders of magnitude larger than that of an isothermal gas considered in PM04. PM04 employ plasma loading on all field lines up to the equator whereas we confine the plasma strictly within a small polar cap. The excess mass loaded onto the field lines beyond the polar cap in PM04 provides additional lateral support which helps in the formation of large mounds and possible eventual burial of the local field.  

In our current work, we have not addressed the burial problem directly, as we have limited ourselves to accretion mounds strictly confined within the polar cap with a fixed boundary far from the surface. Such a situation is more applicable to HMXB or young LMXB systems where higher field strengths and lower accretion rates result in mounds of smaller mass and size. If such a neutron star is to evolve into a low-field object by field burial, the process should commence at this early stage itself.  However, our current work hints that pressure driven instabilities, which have so far been ignored, may play an important role in limiting the efficacy of the burial process. 

\begin{figure}
	\centering
        \includegraphics[width = 6.cm, height = 9.cm,keepaspectratio] {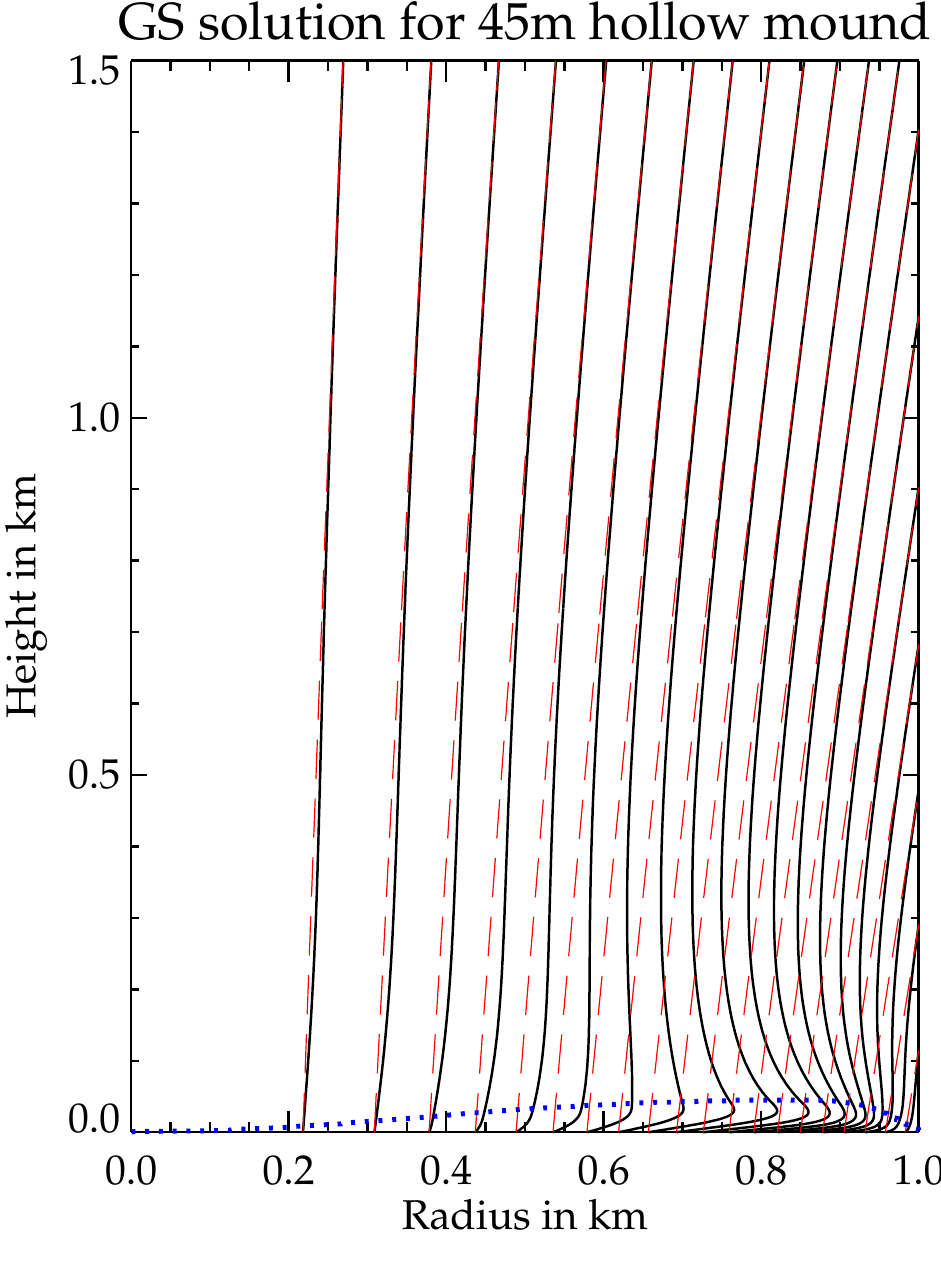}
	\caption{\small Field lines from the GS solution for a hollow mound (in black) of height $\sim 45$m and  pre-accretion dipolar field (in red). The blue dotted line represents the top of the mound. There is significant distortion in the field from the initial dipole nature for several hundred metres above mound ($\sim 5\%$ deviation from dipole field at $\sim 500$m). Far from the neutron star surface ($\sim 1$km and above) the field lines become dipolar.} \label{45mGSextended}
\end{figure}
\begin{figure}
	\centering
        \includegraphics[width = 7.cm, height = 7.cm,keepaspectratio] {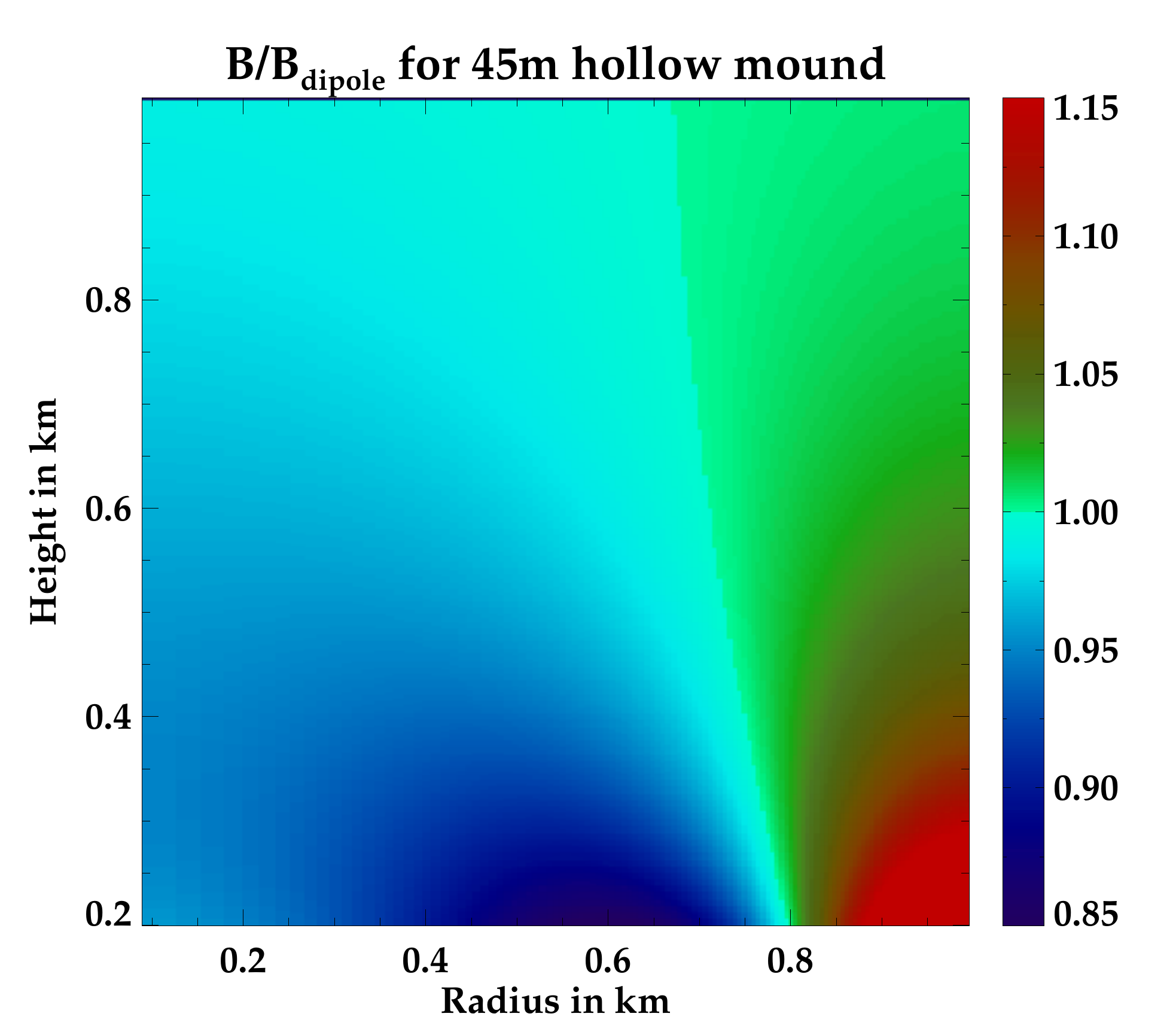}
	\caption{\small Ratio of $B_{\rm GS}$ to $B_{\rm dipole}$ for the solution in Fig.~\ref{45mGSextended}. The field is plotted for heights $\geq 200$m to represent the field variation in the column above the mound. Field lines pushed to the periphery cause enhancement of magnetic field at the outer wall of the column, and decrease in strength at the inner walls.} \label{fieldcompare}
\end{figure}
\begin{figure}
	\centering
        \includegraphics[width = 7.cm, height = 7.cm,keepaspectratio] {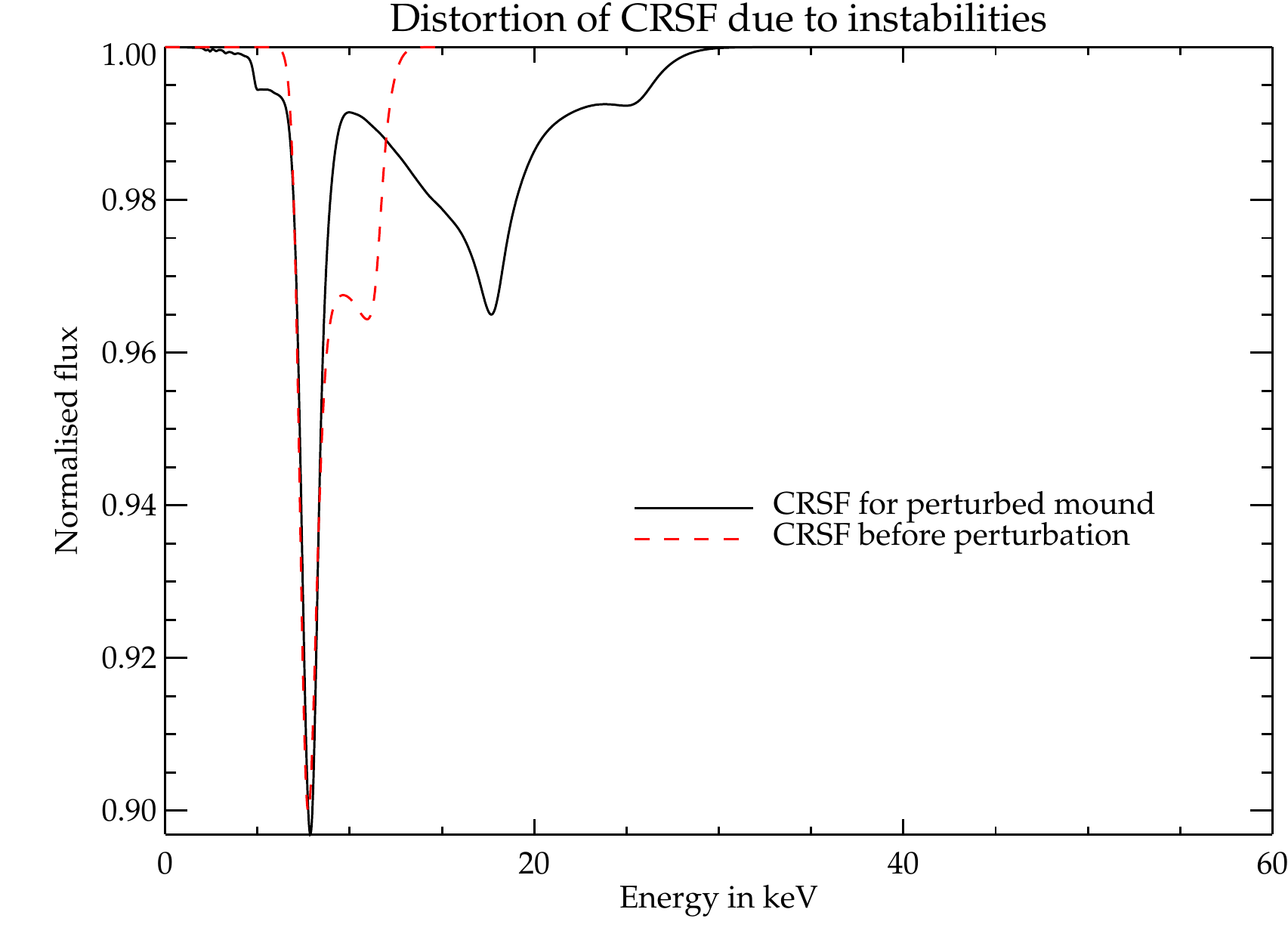}
	\caption{\small  CRSF calculated for the field at the top of a PLUTO domain for a $Z_c=50$m mound. The red line shows the CRSF before the onset of instabilities. Distortion in the field due to instabilities broaden the CRSF. In reality the CRSF profile may be different due to reprocessing of the profile in the accretion column, but there will be a general tendency of broadening of the CRSF with the onset of instabilities. } \label{pluto50mcrsf}
\end{figure}
Since our current PLUTO simulations are limited to a vertical extent less than the mound height, dynamic field distortions above the mound cannot be probed by them. However, from the static GS solutions in Fig.~\ref{45mGSextended} and Fig.~\ref{fieldcompare}, it is clear that although the mound is only a few tens of metres high, the distortions in the field lines last for several hundred metres.  This implies that distortions in the mound structure due to the MHD instabilities will also affect the field structure far away from the mound, leaving their imprint on the cyclotron lines (CRSF) from the accretion column.  

Any distortion in the field lines will introduce broadening and complexity in the CRSF structure. To illustrate this we have evaluated the CRSF by taking the field distribution at the top of the PLUTO domain at the end of the simulation run ($\sim 3.6 t_A$) of a $Z_c=50$m mound using the method outlined in MB12\footnote{It is to be noted that this exercise may not reproduce realistic CRSF, as the X-ray emission may be further reprocessed in the overlying accretion column.}. Fig.~\ref{pluto50mcrsf} shows the effect of broadening of the CRSF due to superposition of spectra from different parts of the mound. The extent of such broadening is expected to increase with accretion rate.   

\begin{figure}
	\centering
        \includegraphics[width = 7.cm, height = 7.cm,keepaspectratio] {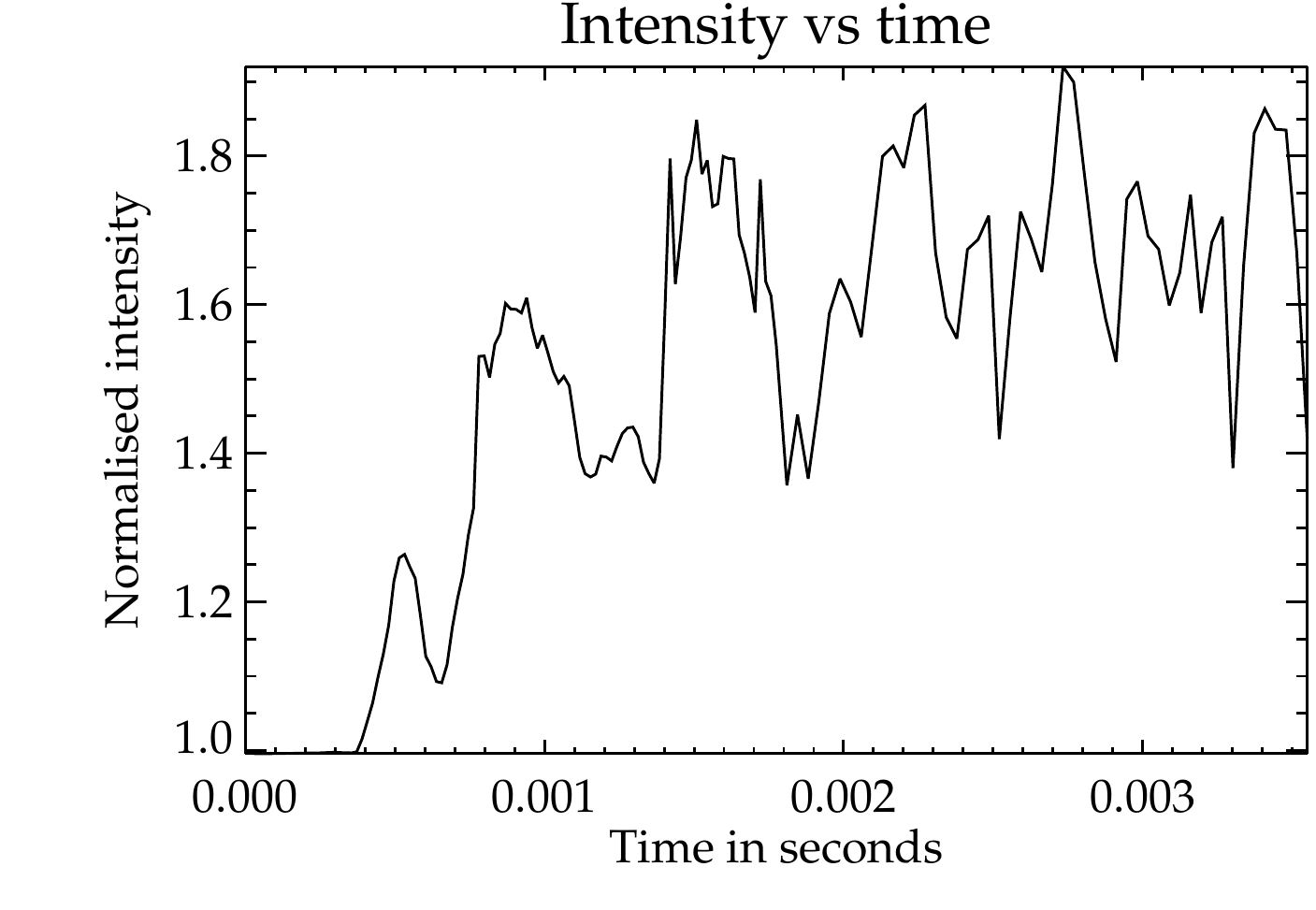}
        \includegraphics[width = 7.cm, height = 7.cm,keepaspectratio] {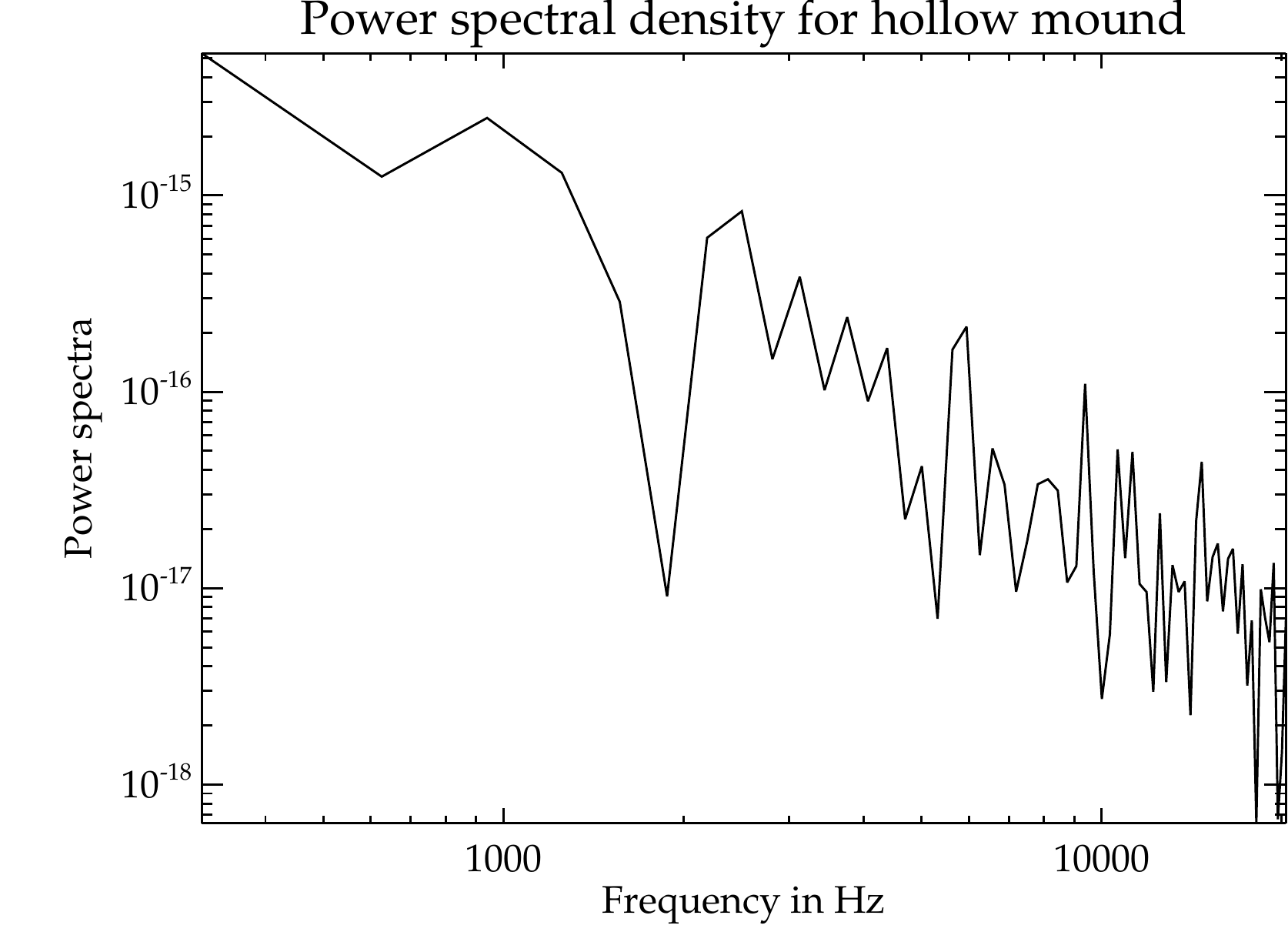}
	\caption{\small Top: Intensity of emission from the top of the mound assuming $I\propto \rho ^2$. Bottom: Power spectral density of the lightcurve.} \label{holopds}
\end{figure}
MHD instabilities may also produce observable timing signatures from accreting neutron stars. Although any radiation from the mound will be reprocessed in the overlying column, oscillations set up in the mound will be carried higher up in the column by the magnetic field, which will add to the noise observed in the power spectra from these systems. The rapid formation of the density streams within  a few Alfv\'en times and subsequent oscillations in similar time scales will inject power in the fluctuation spectrum at frequencies corresponding to local instability growth rates. We have calculated the power spectral density from the Fourier transform of a simulated light curve from the surface of a  hollow mound of $Z_c=45$m, by assuming the intensity of the local emission to depend on density as $I \propto \rho ^2$ (see Fig.~\ref{holopds}). We integrate this over the top of the PLUTO domain to get the flux as a function of time, for a duration of $10 t_A= 3.5\times10^{-3}$s, corresponding to the longest run amongst the current set of simulations. The lightcurve clearly shows repetitive patterns which contribute to a broad feature of excess power around $\sim 800$Hz in the power spectrum. Bumps at kilo Hertz frequencies corresponding to local Alfv\'en time scales are also seen. Thus MHD instabilities can result in oscillations in the emergent intensity which can show up as broad features in the power spectra of observed X-ray flux, and also contribute to high frequency noise.

Although power spectra of HMXB systems are mostly dominated by low frequency features arising in the accretion disc, some sources such as Cen X-3 \citep{jernigan2000} show excess power in the high frequency regime which has been attributed to instabilities in the accretion column. Some sources show correlation between the photon index and the frequency of the main noise component which indicates that they may originate from the same physical region inside the accretion column \citep{reig13}. For mounds of smaller mass, the growth rates being smaller, the characteristic frequencies at which the power would peak due to the MHD processes would also be lower. However, to  explore the nature of the power spectra at lower frequencies, longer simulation runs need to be performed.    

\section{Acknowledgements}
We thank CSIR India for Junior Research Fellow grant, award no 09/545(0034)/2009-EMR-I. We thank Dr. Petros Tzeferacos for his help and suggestions in setting up the boundary conditions in the PLUTO simulations. We also thank Dr. Kandaswamy Subramanian of IUCAA and Dr. Biswajit Paul of RRI for useful discussions and suggestions during the work, and IUCAA HPC team for their help in using the IUCAA HPC where most of the numerical computations were carried out. We thank the anonymous referee for his/her kind comments. DB acknowledges the hospitality of ISSI, Berne and discussions with the Magnet collaboration which have benefited the paper.

%==========================================
%----------------- Bibliography and bibfile
\def\apj{ApJ}%
\def\mnras{MNRAS}%
\def\aap{A\&A}%
\def\apjl{ApJ}
\def\physrep{PhR}
\def\apjs{ApJS}
\def\pasa{PASA}
\def\pasj{PASJ}
\def\nat{Nature}
\def\memsai{MmSAI}

\bibliographystyle{mn2e}
\bibliography{dipanjanbib}

\end{document}